\documentclass{article}

\usepackage{PRIMEarxiv}

\usepackage[utf8]{inputenc} %
\usepackage[T1]{fontenc}    %
\usepackage{hyperref}       %
\usepackage{url}            %
\usepackage{booktabs}       %
\usepackage{amsfonts}       %
\usepackage{nicefrac}       %
\usepackage{microtype}      %
\usepackage{fancyhdr}       %
\usepackage{graphicx}       %

\usepackage{natbib}
\usepackage{amsmath}
\usepackage{amssymb}
\usepackage{amsbsy}
\usepackage{amsthm}
\usepackage{mathtools}
\usepackage{tikz}
\usepackage{bbm}
\usepackage[caption=false]{subfig}

\DeclareMathOperator*{\argmax}{arg\,max}
\newcommand{\matpower}[0]{{\sc Matpower}}

\graphicspath{{media/}}     %

\title{Network Cascade Vulnerability using Constrained Bayesian Optimization
}

\author{
  Albert Lam \\
  Argonne National Laboratory \\
  \And 
  Mihai Anitescu \\
  Argonne National Laboratory \\
  \And
  Anirudh Subramanyam \\
  The Pennsylvania State University \\
}

\begin{document}
\maketitle

\begin{abstract}
Measures of power grid vulnerability are often assessed by the amount of damage an adversary can exact on the network. 
However, the cascading impact of such attacks is often overlooked, even though cascades are one of the primary causes of large-scale blackouts.
This paper explores modifications of transmission line protection settings as candidates for adversarial attacks, which can remain undetectable as long as the network equilibrium state remains unaltered.
This forms the basis of a black-box function in a Bayesian optimization procedure, where the objective is to find protection settings that maximize network degradation due to cascading.
Notably, our proposed method is agnostic to the choice of the cascade simulator and its underlying assumptions.
Numerical experiments reveal that, against conventional wisdom, maximally misconfiguring the protection settings of all network lines does not cause the most cascading.
More surprisingly, even when the degree of misconfiguration is limited due to resource constraints, it is still possible to find settings that produce cascades comparable in severity to instances where there are no resource constraints.
\end{abstract}

\section{Introduction}

Vulnerability analysis of power systems seeks to identify weaknesses in the electrical grid.
Within the setting of cyberattacks, this usually postulates an adversary that attempts to devise malicious attacks on the network's cyber-physical layer that lead to network degradation.
These are often formulated as optimization problems, where the goal is to maximize the amount of damage that can be inflicted on the network after the adversary gains access to parts of the grid's cyber-physical infrastructure.
Crucially, improving the grid's resilience to these worst-case scenario attacks depends on how well the formulation of these problems mirror the realities of such scenarios.

Cyber-physical attacks often improve upon the effectiveness of ordinary physical attacks by obscuring changes in the network state or topology via cyber intrusions of the grid's Supervisory Control and Data Acquisition (SCADA) system~\citep{zhang2016cyberphys,kumar2021detectstealthy}.
These can elevate to become undetectable attacks, which aim to prolong the time before operators uncover the intrusion and intervene to make a correction, by preserving the grid's equilibrium state through maintaining power flows across each line and power balances at each bus, in addition to abiding by the safety limits on the grid's voltages, angles, and line flows. 
Measurements or estimations of the network's state would then appear to indicate that the grid is operating safely.

In this paper, we study undetectable attacks on transmission lines specifically.
Although these can take many forms~\citep{he2016cyber}, we focus on the idea of tampering with digital protective relays, which are critical elements of modern power systems that protect transmission lines from faults.
It is widely recognized that the communication systems and protection settings of digital relays are prime targets for adversaries to hijack~\citep{gurevich2014cyber,hong2019resilientdistanceprotection,mcdermott2020cybersecurity,rajkumar2020cyberattacks,amin2020cyber}, since misconfigured relays can produce unintended trip signals that can eventually trigger cascading failures~\citep{pourbeik2006anatomypowergridblackout,baldick2009vulnerabilityassessment}.
The December~2016 cyberattack on the Ukrainian power system exemplifies how tampering of the grid's digital protection devices can cause widespread disruptions~\citep{slowik2019crashoverride}.

Cyberattacks on protective relays are diverse and include, besides others, denial-of-service attacks between the relay and its circuit breaker, and switching attacks~\citep{stefanov2012cyber,gurevich2014cyber,ten2017impact}.
However, we focus on the relatively unexplored domain of data integrity attacks~\citep{rahman2016multi,amin2020cyber}, where the control settings or protection algorithms of the relay are tampered with, which can be done remotely via unsecured or compromised wireless communication channels between the system operator and the protection device.

This phenomenon is best illustrated by considering a simple overcurrent relay.
Here, a trip signal is triggered whenever the measured line current exceeds the relay's fault detection threshold limit, also known as its pickup value.
Lower threshold limits will significantly increase the likelihood of relay tripping, as even small stochastic fluctuations or disturbances in other parts of the grid can cause the line current to exceed the reduced threshold limit, thereby causing a trip.
An adversary is therefore able to increase the likelihood of cascading failures by deliberately reducing these threshold limits, which can also lead to unstable system conditions, since a compromised relay could change the grid topology.
Importantly, the attack will be undetectable as long as the reduced threshold limit is still sufficiently above the equilibrium current flow on the line, which ensures that the relay does not cause a trip during normal steady state conditions.

Analogous control setting misconfigurations are possible for other types of relays as well, as long as they have some inherent digital logic processing capability.
For instance, microprocessor-based distance relays work by tripping circuit breakers when--as a proxy for the occurrence of a fault--sufficiently low apparent impedance is detected on the line~\citep{leelaruji2011protectiverelaying}.
Internally, this is implemented using several zones of protection, each of which is calculated as a fixed predetermined percentage of line impedance.
When the power load on the line increases, the apparent impedance decreases and can potentially ``encroach'' on the relay's protection characteristic.
In the worst case, the load can be so high as to fall within the relay's most conservative fault detection impedance zone, known as Zone~3.
In effect, the relay would misidentify this condition as a fault and trip the line out of service~\citep{rincon2012loadabilitylimits}.
An attacker can take advantage of this load encroachment phenomenon by simply increasing the maximum impedance reach of the relay's Zone~3 setting~\citep{hong2019resilientdistanceprotection}.
As in the case of overcurrent relays, this will increase the likelihood of small upward stochastic fluctuations in load flow to cause an unwanted line trip, and trigger a cascading failure.

We propose that modifying a relay's fault detection threshold--either directly as in overcurrent relays, or indirectly as in distance relays--can be exploited by adversaries as a mechanism to intentionally trigger cascading outages.
These can be classed as malicious ``configuration changes''~\citep{hong2019resilientdistanceprotection} or ``hidden failures''~\citep{elizondo2001hiddenfailures}, but we refer to these modifications as line limit tightening attacks to emphasize their direct impact on line flow limits.
As long as the tightening does not alter the system's equilibrium state, these only become apparent when other random and unpredictable events trigger the relays to malfunction, which we refer to as a set of initial contingencies.
Grid operators then have at most a few minutes to readjust these limits before sequential relay tripping causes widespread cascading failures.

We examine the cascading impact of line limit tightening through the lens of Bayesian optimization (BO), where the black-box objective function is a cascade severity metric computed from the outputs of a cascade simulator.
Each configuration of the network's line limits then corresponds to one proposed sample from which cascades are simulated from before being evaluated under the chosen cascade metric.
The goal of the adversary is to find a configuration that maximizes the cascade metric, while also remaining undetectable until a set of initial contingencies triggers cascading.

We also incorporate sparsity-inducing resource constraints in this BO problem, mirroring existing studies in the vulnerability analysis literature~\citep{zhang2016cyberphys,salmeron2004analysiselectric,kim2016vulnerability}, and refer to these variations collectively as constrained BO problems.
The inclusion of resource constraints have typically been justified based on implicit assumptions that some parts of the grid are less secure than others, such that any prospective adversary only has the capacity to target a sparse subset of the entire grid.
However, we argue such constraints are also important because any prospective adversary would prefer configuration changes that tighten fewer lines as long as the resulting attack can produce an equally devastating cascade, since we do not expect preservation of the equilibrium state to be the sole security measure embedded within power systems in reality.
Furthermore, this is especially relevant in cases where the optimal attack configuration is sufficiently sparse to help focus the efforts of system operators to either secure the corresponding set of protection relays, or prioritize those lines within an $N-k$ contingency ranking scheme in order to satisfy requirements laid out by the North American Electric Reliability Corporation~\citep{nercresiliencystandard}.
In this context, our proposed BO approach can also be viewed as a ``continuous'' extension of the ideas presented in~\citet{gjorgiev2022vulnerabilities}, which seeks to identify sets of contingencies that maximize the load shed from cascading.
The main difference in our work is that instead of searching over the discrete, and combinatorially increasing, space of line contingency subsets, we search over the continuous space of line limits, allowing for a more tractable optimization problem.
The constrained BO problem is then solved via two different avenues: the first relies on trust regions to constrain the maximization of the acquisition function to the feasible region of the search space, and the second uses an exact penalty function to encode penalties for constraint violations directly in the objective function.

Beyond tractability, the most distinguishing feature of our proposed framework is its modularity at each stage of the  modeling process.
This is especially vital in the analysis of cascading failures given the notorious complexity involved in modeling and simulating the plethora of mechanisms by which failures can occur~\citep{baldick2008reviewmethodscascading}.
As generic BO does not assume underlying structure in the objective, it allows for considerable flexibility in the choice of the cascade simulator, including any of the commercially available or research-grade simulators surveyed in~\citet{papic2011surveytoolscascading}, and~\citet{henneaux2018benchmarking}.
The choice made here can then influence several other modeling choices, including the degree of stochasticity for initiating cascading events (e.g., whether electrical variables exceeding their safety thresholds are deterministic or probabilistic); the type of methodology (e.g., dynamic, quasi-steady state, data-driven, etc.); the power flow model (e.g., AC vs DC, and standard vs optimal power flow); and the mechanisms of cascading besides line trips (e.g., voltage violations, frequency instabilities, etc.).
The importance of modularity has also been noted in~\citet{henneaux2018benchmarking}, which highlighted how slight differences in modeling assumptions can simulate vastly different cascades.
The most meaningful output of such simulators then is the overall distribution of the produced samples and their corresponding statistics, which we capture via a cascade severity metric (e.g., expected load shed, number of failed lines, etc.).
Finally, BO itself is highly modular, with subroutines that could be replaced with various alternatives that better reflect the specifics of the problem at hand (e.g., acquisition function, kernel function, etc.).

In addition to developing a BO framework for network vulnerability analysis, we note the following contributions and findings of this paper based on the analysis of a small test system:

\begin{enumerate}  
    \item Unconstrained BO can substantially outperform random sampling of the line limit space, and more surprisingly, finds configurations that produce more severe cascades than the maximal tightening attack in which all line limits are maximally misconfigured.
    \item Even in the constrained setting, BO can find attacks that deal the vast majority of the cascade damage of the best unconstrained attack.
    \item Furthermore, these constrained attacks can be highly sparse, suggesting that purely restricting the amount of tightening is not necessarily a sufficient security measure for preventing such attacks.
\end{enumerate}

The rest of the paper is organized as follows: Section~\ref{sec:network_model} describes the network model, including the process for simulating cascades; Section~\ref{sec:attack_model} describes the BO attack model for how an adversary optimizes for cascade severity; Section~\ref{sec:experiments} presents the results of our experiments; Section~\ref{sec:future_work} discusses future work directions; and Section~\ref{sec:conclusion} concludes the paper.

\section{Network Model}\label{sec:network_model}

\subsection{Power System Preliminaries}\label{sec:network_model:preliminaries}

This section introduces basic notation and terminology commonly used in power systems planning models.
Let $\mathcal{N}$ denote the set of $\lvert \mathcal{N} \rvert$ buses (nodes), and $\mathcal{E} \subseteq \mathcal{N} \times \mathcal{N}$ denote the set of $\lvert \mathcal{E} \rvert$ transmission lines (edges), such that there is a line $(i,j) \in \mathcal{E}$ from bus $i \in \mathcal{N}$ to bus $j \in \mathcal{N}$ if $j \in \mathcal{N}_i$ is a neighbor to $i$.
The complex power flow across line $(i,j) \in \mathcal{E}$ is denoted by $S_{i,j} = p_{i,j} + \mathrm{j} \cdot q_{i,j}$, which has active power $p_{i,j}$ and reactive power $q_{i,j}$ components, and $\mathrm{j}=\sqrt{-1}$ denotes the imaginary unit.
The system admittance matrix is given by $Y =  G + \mathrm{j} \cdot B \in \mathbb{C}^{\lvert \mathcal{N} \rvert \times \lvert \mathcal{N} \rvert}$, and the vector of voltages is denoted by $V \in \mathbb{C}^{\lvert \mathcal{N} \rvert}$, where $V_i = v_i \exp \left(\mathrm{j} \cdot \delta_i \right)$ for all $i \in \mathcal{N}$. We also denote the conjugate of $z \in \mathbb{C}$ by $z^{\ast}$.

Let $\mathcal{L}_i$ be the set of loads attached to bus $i \in \mathcal{N}$, and $\mathcal{G}_i$ be the set of generators attached to bus $i$, so that the set of all loads is $\mathcal{L} = \cup_{i \in \mathcal{N}} \mathcal{L}_i$, and the set of all generators is $\mathcal{G} = \cup_{i \in \mathcal{N}} \mathcal{G}_{i}$.
The power generated by generator $g \in \mathcal{G}_i$ is $S_g^{G} = p_g^G + \mathrm{j} \cdot q_g^G$, and the power demanded by load $l \in \mathcal{L}_i$ is $S_l^{L} = p_l^L + \mathrm{j} \cdot q_l^L$.

For given values of the voltages, the magnitude of the current flow along line $(i,j) \in \mathcal{E}$ from bus~$i$ to bus~$j$ is denoted by $\Theta_{(i,j)}$, and can be computed via $\Theta_{(i,j)} = \left \lvert Y_{i,j} (V_i - V_j) \right \rvert$. 
System operators then seek to control the power generations $S^G$ and voltages $V$ so that the induced current flow $\Theta_{(i,j)}$ along each line $(i,j) \in \mathcal{E}$ remains below a maximum line limit, which we denote by $\Theta_{(i,j)}^{\max}$.
In practice, $\Theta_{(i,j)}^{\max}$ is typically smaller than the fault detection threshold beyond which protection relays would automatically disconnect that line.
For simplicity of notation, however, our exposition assumes that these threshold values are equal, so that line $(i,j)$ fails irrecoverably if the flow $\Theta_{(i,j)}$ exceeds $\Theta_{(i,j)}^{\max}$.

\subsection{Equilibrium State}\label{sec:network_model:equilbrium}

Prior to any attack or line failure, we assume that the network operates in a steady state equilibrium that satisfies all power balances and flow limits.
Although there is some variety here, operators most commonly achieve this by computing a solution to the alternating current optimal power flow (ACOPF) problem:
\begin{align}
    \mathop{\text{minimize}}_{V, S^G}\;  & \sum_{g \in \mathcal{G}} c_g(p_g^G) \label{eqn:acopf_obj} \qquad & \\
    \text{subject to} \; & \sum_{g \in \mathcal{G}_i} S_g^{G} - \sum_{l \in \mathcal{L}_i} S_l^{L} = \sum_{j \in \mathcal{N}_i} Y_{i,j}^{\ast} \left(\lvert V_i \rvert^2 - V_i V_j^{\ast} \right) \qquad &\forall i \in \mathcal{N}, \forall (i,j) \in \mathcal{E} \label{eqn:power_balance_and_power_flow} \\
    & V_i^{\min} \le V_i \le V_i^{\max}, p_g^{G, \min} \le p_g^G \le p_g^{G, \max}, q_g^{G, \min} \le q_g^G \le q_g^{G, \max} \qquad &\forall i \in \mathcal{N}, \forall g \in \mathcal{G} \label{eqn:voltage_power_gen_limits} \\
    & \Theta_{(i,j)}\le \Theta_{(i,j)}^{\max} \qquad &\forall (i,j) \in \mathcal{E}. \label{eqn:power_flow_limits}
\end{align}

The objective function~\eqref{eqn:acopf_obj} seeks to minimize the aggregate cost $c_g$ of generating active power $p_g^{G}$ across all generators, subject to constraints~\eqref{eqn:power_balance_and_power_flow} that enforce Kirchhoff's laws of power balance; and constraints~\eqref{eqn:voltage_power_gen_limits}-\eqref{eqn:power_flow_limits} which ensure that the computed voltages, generated power, and line flows remain within the safety limits of the system.
We underscore that the solution of~\eqref{eqn:acopf_obj}-\eqref{eqn:power_flow_limits} only depends on the network's line limits through constraint~\eqref{eqn:power_flow_limits}.
In particular, if the current flows of the ACOPF solution are $\bar{\Theta}$, and then the line limits (or equivalently, the relays' fault detection thresholds) are modified such that the new limits $\Theta^{x}$ satisfy $\Theta_{(i,j)}^{x} \geq \bar{\Theta}_{(i,j)}$, then the optimal ACOPF solution remains unchanged.

\subsection{Line Limit Tightening Attacks} \label{sec:network_model:line_limit_tightening}

For each line $(i,j) \in \mathcal{E}$, we introduce a continuous tightening parameter $x_{(i,j)} \in [0,1]$ to model the degree with which the line limit $\Theta_{(i,j)}^{\max}$ is reduced.
Specifically, suppose that $\bar{\Theta}$ represents the vector of equilibrium current flows corresponding to an optimal solution of~\eqref{eqn:acopf_obj}-\eqref{eqn:power_flow_limits}, and the tightened line limit $\Theta_{(i,j)}^{x}$ is constrained to be bounded between $\bar{\Theta}_{(i,j)}$ and the original unmodified line limit $\Theta_{(i,j)}^{\max}$, such that:
\begin{equation}
\Theta_{(i,j)}^{x} = x_{(i,j)} \bar{\Theta}_{(i,j)} + \left(1 - x_{(i,j)} \right) \Theta_{(i,j)}^{\max}. \label{eqn:tightened_line_limit_def} 
\end{equation}

Observe that $\Theta_{(i,j)}^{0} = \Theta_{(i,j)}^{\max}$ and $\Theta_{(i,j)}^{1} = \bar{\Theta}_{(i,j)}$, so that larger values of $x_{(i,j)}$ give a tighter line limit, and by construction, any $x_{(i,j)} \in [0,1]$ satisfies $\Theta_{(i,j)}^{x} \ge \bar{\Theta}_{(i,j)}$. 
It follows that once the network is in equilibrium, modifying any line limit according to~\eqref{eqn:tightened_line_limit_def} will not alter the network's state, since this gives the same minimizer from solving~\eqref{eqn:acopf_obj}-\eqref{eqn:power_flow_limits}, and the resulting line limit tightening attack will be undetectable.

\subsection{Network Cascading} \label{sec:network_model:cascading}

Following a line limit tightening attack, random contingency events are more likely to trigger cascading across the network.
Similar to previous work~\citep{subramanyam2022fpacopf}, we employ a probabilistic model of initial contingencies that is commonly used for the simulation of cascading failures in power systems~\citep{henneaux2018benchmarking,zhou2023mostfrequentoutages}, where the network is randomly perturbed by a set of two initial line contingencies, as we conservatively assume it is resilient to single-element contingencies.
Cascade severity is then measured as the ensuing cumulative damage following these contingencies--as previously noted, we can use any valid model for simulating cascades so long as the output can be used to compute sample statistics of cascade severity.
For our experiments, we use the simulator previously introduced in~\citet{roth2021kmc}, known as Kinetic Monte Carlo (KMC).
Broadly, the KMC simulator models the dynamic evolution of the system state, taking into account small stochastic perturbations in generation and demand, that can eventually force the system to exit a ``basin of attraction'' for a particular line.
Using large deviation theory, the ``first exit time'' from this basin is modeled as an exponential random variable with failure rate $\lambda_{(i,j)}$, which can be computed by solving a constrained nonlinear optimization problem.
Within the power systems cascade simulation literature, the KMC simulator can be classed as a quasi-steady state simulator that uses asymptotic approximations for estimating the failure rates that drive cascading failures.
Further assumptions and details regarding the underlying simulator can be found in~\citet{roth2021kmc}.
  
\section{Attack Model}\label{sec:attack_model}

\subsection{Optimizing Attacks} \label{sec:experiments:optimizing_attacks}
The attack model implicitly assumes that the adversary has gained access to the grid's infrastructure, that they are able to observe (or estimate) the equilibrium line flows $\bar{\Theta}$, and are able to launch an attack via line limit tightening as described in Section~\ref{sec:network_model:line_limit_tightening}.
Their goal is to maximize a cascade severity metric, denoted $f(x)$, by choosing a set of tightening parameters $x \coloneqq \{x_{(i,j)}, \hspace{0.5em} (i,j) \in \mathcal{E}\}$ to apply to each line $(i,j) \in \mathcal{E}$, which we refer to as a line limit configuration. 
In our experiments, $f(x)$ is chosen to be the empirically observed average number of line failures across a set of approximately $200$ cascade simulations, which together comprise one evaluation of the severity metric for configuration $x$, and is a commonly used metric of cascade severity~\citep{henneaux2018benchmarking}.
Each simulation is allowed to run until no more lines are available for failure, or until a maximum cascade duration has elapsed, which we set to $10^5$ seconds (roughly $1$ day).

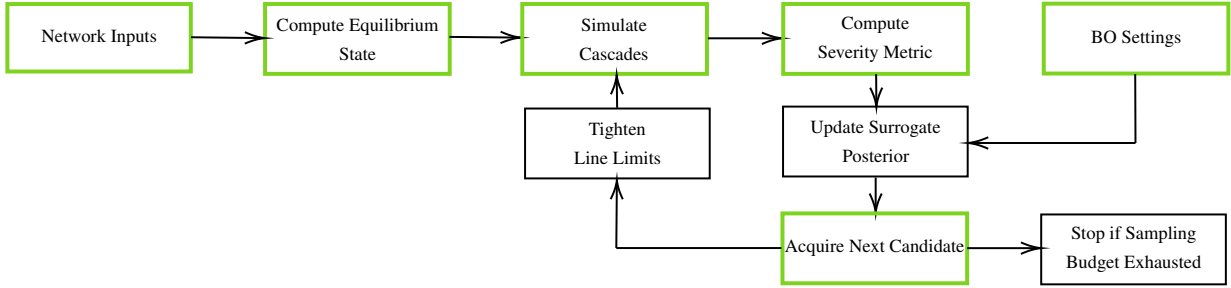
\begin{figure*}[ht]
    \centering
    \captionsetup[subfigure]{justification=centering}

    \tikzset{every picture/.style={line width=0.75pt}}   

    \begin{tikzpicture}[x=0.7pt,y=0.7pt,yscale=-0.95,xscale=1.0]

    \draw  [color={rgb, 255:red, 126; green, 211; blue, 33 }  ,draw opacity=1 ][line width=1.5]  (20.75,11.09) -- (119.87,11.09) -- (119.87,49.5) -- (20.75,49.5) -- cycle ;

    \draw  [color={rgb, 255:red, 126; green, 211; blue, 33 }  ,draw opacity=1 ][line width=1.5]  (160,11.08) -- (259.78,11.08) -- (259.78,50.86) -- (160,50.86) -- cycle ;

    \draw    (120,30) -- (158,30.48) ;
    \draw [shift={(160,30.5)}, rotate = 180.72] [color={rgb, 255:red, 0; green, 0; blue, 0 }  ][line width=0.75]    (10.93,-3.29) .. controls (6.95,-1.4) and (3.31,-0.3) .. (0,0) .. controls (3.31,0.3) and (6.95,1.4) .. (10.93,3.29)   ;
    \draw    (259.6,29.8) -- (296.2,30.18) ;
    \draw [shift={(298.2,30.2)}, rotate = 180.59] [color={rgb, 255:red, 0; green, 0; blue, 0 }  ][line width=0.75]    (10.93,-3.29) .. controls (6.95,-1.4) and (3.31,-0.3) .. (0,0) .. controls (3.31,0.3) and (6.95,1.4) .. (10.93,3.29)   ;
    \draw  [color={rgb, 255:red, 126; green, 211; blue, 33 }  ,draw opacity=1 ][line width=1.5]  (299,11.08) -- (398.78,11.08) -- (398.78,50.86) -- (299,50.86) -- cycle ;

    \draw    (399.4,30.6) -- (437.4,30.6) ;
    \draw [shift={(439.4,30.6)}, rotate = 180] [color={rgb, 255:red, 0; green, 0; blue, 0 }  ][line width=0.75]    (10.93,-3.29) .. controls (6.95,-1.4) and (3.31,-0.3) .. (0,0) .. controls (3.31,0.3) and (6.95,1.4) .. (10.93,3.29)   ;
    \draw  [color={rgb, 255:red, 126; green, 211; blue, 33 }  ,draw opacity=1 ][line width=1.5]  (440,11.08) -- (539.78,11.08) -- (539.78,50.86) -- (440,50.86) -- cycle ;

    \draw    (350.25,70.15) -- (350.03,53.5) ;
    \draw [shift={(350,51.5)}, rotate = 89.23] [color={rgb, 255:red, 0; green, 0; blue, 0 }  ][line width=0.75]    (10.93,-3.29) .. controls (6.95,-1.4) and (3.31,-0.3) .. (0,0) .. controls (3.31,0.3) and (6.95,1.4) .. (10.93,3.29)   ;
    \draw   (300.5,70.58) -- (400.28,70.58) -- (400.28,110.36) -- (300.5,110.36) -- cycle ;

    \draw    (490.2,51) -- (490.33,67.65) ;
    \draw [shift={(490.35,69.65)}, rotate = 269.54] [color={rgb, 255:red, 0; green, 0; blue, 0 }  ][line width=0.75]    (10.93,-3.29) .. controls (6.95,-1.4) and (3.31,-0.3) .. (0,0) .. controls (3.31,0.3) and (6.95,1.4) .. (10.93,3.29)   ;
    \draw   (439.9,69.48) -- (539.68,69.48) -- (539.68,109.26) -- (439.9,109.26) -- cycle ;

    \draw    (489.95,109.9) -- (489.95,127.9) ;
    \draw [shift={(489.95,129.9)}, rotate = 270] [color={rgb, 255:red, 0; green, 0; blue, 0 }  ][line width=0.75]    (10.93,-3.29) .. controls (6.95,-1.4) and (3.31,-0.3) .. (0,0) .. controls (3.31,0.3) and (6.95,1.4) .. (10.93,3.29)   ;
    \draw  [color={rgb, 255:red, 126; green, 211; blue, 33 }  ,draw opacity=1 ][line width=1.5]  (439.6,130.28) -- (539.38,130.28) -- (539.38,170.06) -- (439.6,170.06) -- cycle ;

    \draw    (349.8,150) -- (439,150) ;
    \draw  [color={rgb, 255:red, 126; green, 211; blue, 33 }  ,draw opacity=1 ][line width=1.5]  (580.78,10.88) -- (680.55,10.88) -- (680.55,50.66) -- (580.78,50.66) -- cycle ;

    \draw    (539.38,150.06) -- (577.38,150.06) ;
    \draw [shift={(579.38,150.06)}, rotate = 180] [color={rgb, 255:red, 0; green, 0; blue, 0 }  ][line width=0.75]    (10.93,-3.29) .. controls (6.95,-1.4) and (3.31,-0.3) .. (0,0) .. controls (3.31,0.3) and (6.95,1.4) .. (10.93,3.29)   ;
    \draw    (630.35,90.2) -- (541.85,90.2) ;
    \draw [shift={(539.85,90.2)}, rotate = 360] [color={rgb, 255:red, 0; green, 0; blue, 0 }  ][line width=0.75]    (10.93,-3.29) .. controls (6.95,-1.4) and (3.31,-0.3) .. (0,0) .. controls (3.31,0.3) and (6.95,1.4) .. (10.93,3.29)   ;
    \draw    (630.2,51) -- (630.35,90.2) ;
    \draw   (579.4,130.88) -- (679.18,130.88) -- (679.18,170.66) -- (579.4,170.66) -- cycle ;

    \draw    (349.8,150) -- (350.1,113.27) ;
    \draw [shift={(350.12,111.27)}, rotate = 90.47] [color={rgb, 255:red, 0; green, 0; blue, 0 }  ][line width=0.75]    (10.93,-3.29) .. controls (6.95,-1.4) and (3.31,-0.3) .. (0,0) .. controls (3.31,0.3) and (6.95,1.4) .. (10.93,3.29)   ;

    \draw (350.52,90.34) node   [align=left] {\begin{minipage}[lt]{67.85pt}\setlength\topsep{0pt}
    \begin{center}
    {\scriptsize Tighten }\\{\scriptsize Line Limits}
    \end{center}

    \end{minipage}};
    \draw (348.6,31.29) node   [align=left] {\begin{minipage}[lt]{67.46pt}\setlength\topsep{0pt}
    \begin{center}
    {\scriptsize Simulate }\\{\scriptsize Cascades}
    \end{center}

    \end{minipage}};
    \draw (489.8,31.04) node   [align=left] {\begin{minipage}[lt]{68.54pt}\setlength\topsep{0pt}
    \begin{center}
    {\scriptsize Compute }\\{\scriptsize Severity Metric}
    \end{center}

    \end{minipage}};
    \draw (210.02,31.04) node   [align=left] {\begin{minipage}[lt]{67.85pt}\setlength\topsep{0pt}
    \begin{center}
    {\scriptsize Compute Equilibrium State}
    \end{center}

    \end{minipage}};
    \draw (70.38,30.44) node   [align=left] {\begin{minipage}[lt]{67.49pt}\setlength\topsep{0pt}
    \begin{center}
    {\scriptsize Network Inputs}
    \end{center}

    \end{minipage}};
    \draw (489.92,89.24) node   [align=left] {\begin{minipage}[lt]{67.85pt}\setlength\topsep{0pt}
    \begin{center}
    {\scriptsize Update Surrogate Posterior}
    \end{center}

    \end{minipage}};
    \draw (489.62,150.04) node   [align=left] {\begin{minipage}[lt]{67.85pt}\setlength\topsep{0pt}
    \begin{center}
    {\scriptsize Acquire Next Candidate}
    \end{center}

    \end{minipage}};
    \draw (630.8,30.84) node   [align=left] {\begin{minipage}[lt]{67.85pt}\setlength\topsep{0pt}
    \begin{center}
    {\scriptsize BO Settings}
    \end{center}

    \end{minipage}};
    \draw (629.42,150.64) node   [align=left] {\begin{minipage}[lt]{67.85pt}\setlength\topsep{0pt}
    \begin{center}
    {\scriptsize Stop if Sampling Budget Exhausted}
    \end{center}

    \end{minipage}};

    \end{tikzpicture}

    \caption{Flowchart of how an adversary uses BO to optimize attacks.}
    \label{fig:flowchart}
\end{figure*}

We attempt to find the best empirical configuration using BO, where the overall procedure is summarized in Figure~\ref{fig:flowchart}.
Here, the network is in equilibrium when initial contingencies trigger cascading, subject to the configuration set by the adversary. 
The simulated cascades are then summarized in the severity metric, which is used to update the posterior of the surrogate model before a new configuration is sampled. 
This continues until the sampling budget of the adversary is exhausted. 
As alluded to earlier, each of the stages highlighted in green could be replaced with various alternatives.

\subsection{Unconstrained BO}\label{sec:attack_model:unconstrained_BO}
Given cascade severity metric $f \colon [0,1]^{\lvert \mathcal{E} \rvert} \mapsto \mathbb{R}$, which we want to maximize over line limit configuration candidates $x \in X \coloneqq [0,1]^{\lvert \mathcal{E} \rvert}$, the goal of the unconstrained BO formulation is to solve:
\begin{align}\label{eqn:unconstrained_problem}
    \mathop{\text{maximize}}_{x \in X} \; f(x).
\end{align}

The basic idea of BO~\citep{frazier2018tutorial} is to replace the objective function $f$ with a tractable surrogate model, the most common choice being a Gaussian process $\tilde{f} \sim \mathcal{GP}(m, k)$ with mean function
$m \colon X \mapsto \mathbb{R}$ and covariance kernel $k \colon X \times X \mapsto \mathbb{R}$.
Once we have evaluated $f$ at each of the $n$ candidate points $x^{(1)}, \ldots, x^{(n)}$, define $d_n \in \mathbb{R}^n$ and $m_n \in \mathbb{R}^n$ to be the vectors whose $\ell^\text{th}$ entries are $f(x^{(\ell)})$ and $m(x^{(\ell)})$ respectively; define $k_n : X \mapsto \mathbb{R}^n$ to be the vector-valued function such that for any $x \in X$, the $\ell^\text{th}$ entry of $k_n(x)$ is $k(x, x^{(\ell)})$; and let $K_n \in \mathbb{R}^{n \times n}$ be the matrix whose $\ell^\text{th}$ row vector is $k_n(x^{(\ell)})$.
Then, it is well known that for any $x \in X$, the posterior distribution of $\tilde{f}(x)$ conditioned on the observations $f(x^{(\ell)})$ for $\ell = 1, \ldots, n$ is normal $\mathcal{N}(\mu_n(x), \sigma_n^2(x))$~\citep{rasmussen2005gpml}, where
\begin{align}\label{eqn:posterior_mean_variance}
\mu_n(x) &= m(x) + k_n(x)^\top K_n^{-1} (d_n - m_n) \\
\sigma_n^2(x) &= k(x, x) - k_n(x)^\top K_n^{-1} k_n(x).
\end{align}

The original problem~\eqref{eqn:unconstrained_problem} is then replaced with the search for the next candidate point $x^{(n+1)}$, which is done by maximizing an acquisition function $a_n : X \mapsto \mathbb{R}$ such that $x^{(n+1)} = \argmax_{x \in X} a_n(x)$.
We employ the expected improvement acquisition function $a_n(x) = \mathbb{E}\left[\max \{\tilde{f}(x) - d_n^{\ast}, 0 \} \right]$, where the expectation is taken with respect to the posterior distribution $\tilde{f}(x) \mid \{(x_\ell, f(x_\ell)), \ell = 1, \dots, n \} \sim \mathcal{N}(\mu_n(x), \sigma_n^2(x))$ and $d_n^{\ast}$ is the current best objective value (i.e., the largest component of $d_n$), which has the following closed-form expression~\citep{jones2001globaloptmethods}:
\begin{equation}\label{eqn:expected_improvement_gp}
	a_n(x) =\sigma_n(x) \phi\left( z_n(x) \right) + \left(\mu_n(x) - d_n^{\ast} \right) \Phi\left( z_n(x) \right),
\end{equation}
where $z_n(x) = \frac{\mu_n(x) - d_n^{\ast}}{\sigma_n(x)}$, and $\Phi$, $\phi$ are the standard normal cumulative distribution and density functions respectively.
It can also be verified that if $m$ and $k$ have closed-form gradients, then the gradients of $a_n$ can be computed in closed-form, allowing the use of gradient-based optimization routines to solve
$x^{(n+1)} = \argmax_{x \in X} a_n(x)$.
We stop evaluating new candidates when a sampling budget is reached, and declare the candidate $x^{\ast}$ with the largest $f(x^{\ast})$ to be the empirical maximizer of~\eqref{eqn:unconstrained_problem}.

\subsection{Constrained BO}\label{sec:attack_model:constrained_BO}
Define the constraint function $c \colon X \mapsto \mathbb{R}$ such that $c(x) = \| x \|_1 - \lambda$ for any $x \in X$. Here, $\lambda > 0$ is a user-specified hyperparameter such that $x$ is feasible if $\| x \|_1 \le \lambda$ and infeasible otherwise.
This constraint is motivated by the sparsity-inducing properties of the $\ell_1$-norm, which encourages the selection of line limit configurations with a small number of nonzero elements by solving:
\begin{align}
    \mathop{\text{maximize}}_{x \in X} \; f(x) \text{ subject to } & c(x) \le 0. \label{eqn:constrained_problem}
\end{align}
We explore two approaches to handle constraints in BO.
The first, which we call constrained acquisition BO (CABO), uses a trust region based interior point method~\citep{byrd1999interiorpointalg} to restrict optimization of the acquisition function to regions that are feasible, whereas the second, which we call penalized objective BO (POBO), encodes penalties for constraint violations directly in the objective so that infeasible samples are less likely to be selected.

\subsection{Constrained Acquisition BO (CABO)}\label{sec:attack_model:CABO}
Our primary motivation for proposing CABO stems from the fact that the acquisition function in classical BO forumulations are typically optimized using unconstrained gradient-based methods, which are free to select candidate points outside of the feasible region of the search space.
This can be problematic in constrained BO, especially when the feasible region is small, since it can lead to a substantial portion of the sampling budget being used to evaluate infeasible points.

Popular approaches to this problem within the constrained BO literature include modeling the constraints as their own Gaussian processes~\citep{gardner2014bayesian,letham2019noisyconstrainedBO} and then penalizing the acquisition values by the probability that the candidate point is infeasible; or incorporating the constraints into the objective using augmented Lagrangian techniques~\citep{gramacy2016augmentedlagrangian}, however these are only suitable in cases where the constraints are expensive to evaluate or purely black-box functions.

In contrast, our $\ell_1$-norm constraint can be easily evaluated, so that maximizing the acquisition function subject to this constraint can be achieved by traditional constrained optimization techniques.
Specifically, we opt for a trust region based interior point method that utilizes an approximate sequential quadratic programming step with trust regions for each iteration of a barrier formulation of the constrained optimization problem~\citep{byrd1999interiorpointalg}.
We use this to replace the usual L-BFGS-B algorithm used to maximize the acquisition in our other BO formulations, while the rest of the BO routine is identical to unconstrained BO described above.

\subsection{Penalized Objective BO (POBO)}\label{sec:attack_model:POBO}
Rather than constraining the acquisition function, we also investigate penalizing constraint violations directly in the objective function.
In particular, POBO solves the following unconstrained BO problem:
\begin{equation}\label{eqn:POBO}
	\mathop{\text{maximize}}_{x \in X} \; f(x) + \rho \min \left\{-c(x), 0 \right\},
\end{equation}
where $\rho > 0$ is a fixed scalar penalty parameter.
Observe that the objective function of POBO is identical to that of~\eqref{eqn:constrained_problem} in regions where $c(x) \leq 0$, but is strictly smaller in infeasible regions where $c(x) > 0$.
Problem~\eqref{eqn:POBO} can be viewed as a regularization of the BO objective~\citep{liu2023sparseBO}, although the penalty term specifies the level of desired sparsity, instead of trying to maximize it as part of a multi-objective optimization problem. 
In that setting, combining multiple objectives into a single objective function is commonly achieved via scalarization~\citep{marler2010weightedsum}.
Although our approach is similar in spirit, our problem formulation is firmly in the single objective domain. 

Under mild regularity conditions, it is well known~\citep{han1979exact} that the above penalty function is exact.
Specifically, every local solution of the constrained problem~\eqref{eqn:constrained_problem} is also a local solution of~\eqref{eqn:POBO} whenever $\rho$ is chosen to be larger than the optimal Lagrange multiplier of~\eqref{eqn:constrained_problem}.
Like CABO, POBO is identical to unconstrained BO, except that the vector of prior objective evaluations must be modified so that the $\ell^\text{th}$ element of the vector $d_n$ corresponding to $x^{(\ell)}$ is equal to $f(x^{(\ell)}) + \rho \min\{-c(x^{(\ell)}), 0\}$. 

\subsection{Implementation} \label{sec:attack_model:implementation}
Our experiments are implemented in DeepHyper~\citep{balaprakash2018deephyper}, a scalable hyperparameter search package built on top of Python's scikit-optimize package~\citep{head2021scikitoptimize} that provides a generic and flexible BO routine.
Here, the key idea is to recognize $x$ as a set of hyperparameters in a black-box search problem that optimizes for cascade severity.
In contrast to classical BO in which candidate configurations are sequentially evaluated one at a time, DeepHyper supports simultaneous evaluations of multiple candidates in parallel using a constant liar strategy.
Notably, this allows us to use distributed parallel computing resources to obtain an order of magnitude improvement in computational time, which is hugely beneficial since each evaluation of $f(x)$ can be particularly expensive.

Although there are a range of hyperparameters that can be tuned during BO, we opt for the simplest approach where the covariance and acquisition functions are fixed, and focus only on finding optimal line limit configurations.
For the covariance kernel, we chose a Mat\'ern $3/2$ kernel based on the cross-validated validation error of $512$ randomly sampled evaluations of the objective; the additional noise variance when fitting the Gaussian process is set to $0.5$; and we optimize the acquisition functions using either the trust region based constrained algorithm for CABO, or L-BFGS-B for all other BO routines, where the improvement over the previous best value is set to $0.01$ and the number of optimizer iterations is capped at $2000$.
 
\section{Experiments} \label{sec:experiments}

\subsection{Setup} \label{sec:experiments:setup}
Our experiments are conducted on the IEEE \matpower~30-bus test case, which contains $41$ transmission lines~\citep{zimmerman2011matpower}. 
For simplicity, we set all network and simulator-related parameters in line with previous work using KMC~\citep{roth2021kmc,subramanyam2022fpacopf}. 
Using the hyperparameter settings described in Section~\ref{sec:attack_model:implementation}, we perform three sets of BO experiments using unconstrained BO, CABO, and POBO.
For CABO and POBO, we assume that the adversary is restricted to a fixed budget of tightening where $\lambda = 15$, roughly $37\%$ of the maximum possible tightening for this test case.
We compare these results with several baselines that randomly sample $x_{(i,j)}$, as well as enforce minimal tightening $x^{\min} \coloneqq \{x_{(i,j)} = 0, (i,j) \in \mathcal{E}\}$ and maximal tightening $x^{\max} \coloneqq\{x_{(i,j)} = 1, (i,j) \in \mathcal{E}\}$.

\subsection{Best empirical attacks} \label{sec:results:best_empirical_attacks}
Table~\ref{table:best_empirical_attacks} summarizes the results from our experiments in finding attacks that cause the most cascade damage, as measured by the average number of line failures across approximately $200$ cascade simulations.
Here, $x^{\ast}$ denotes the best attack found within the budget of evaluations, while $f(x^{\ast})$ is the objective function evaluated at $x^{\ast}$.
For the CABO and POBO results, the best attack must also be feasible (i.e. $\|x^{\ast}\|_1 \le \lambda = 15$).
It follows that the POBO results are unpenalized values and can be compared directly with the other experiments that naturally do not have penalties.

For the random sampling experiment, we perform $512$ evaluations and compare this with the BO experiments, where we report our results across two phases: $256$ initial evaluations where $x$ is randomly sampled, followed by $256$ evaluations where $x$ is found via the corresponding BO routine. 
For the constrained problem, we compare constrained BO with a similar randomly sampled experiment, where the sampled tightening parameters are re-scaled to sum to $\lambda = 15$, and denote this as feasible random sampling.
Finally, since all evaluations of the minimal and maximal tightening baselines correspond to the same line limit configuration, we simply report a single evaluation of each in the first two columns.

\begin{table}[htb]
    \footnotesize
    \caption{Best found empirical attacks across all experiments.}
    \label{table:best_empirical_attacks}
    \centering
    \begin{tabular}{@{\hspace*{2.0em}}l@{\hspace*{14.0em}}cccc@{}}
        \toprule
        & \multicolumn{2}{c}{Initial evaluations} & \multicolumn{2}{c}{BO evaluations} \\
        \midrule
        \multicolumn{1}{l}{Experiment} & $ \|x^{\ast}\|_1 $ & $f(x^{\ast})$ & $\|x^{\ast}\|_1$ & $f(x^{\ast})$ \\
        \midrule \midrule
        \multicolumn{1}{l}{Random sampling}             & $23.0$  &   $20.6$  &  \textemdash  &   \textemdash   \\
        \multicolumn{1}{l}{Unconstrained BO}            & $24.1$  &   $20.8$  &  $27.6$       &   $23.0$        \\
        \multicolumn{1}{l}{Feasible random sampling}    & $15.0$  &   $17.2$  &  \textemdash  &   \textemdash   \\
        \multicolumn{1}{l}{CABO}                        & $14.8$  &   $15.6$  &  $14.9$       &   $15.5$        \\
        \multicolumn{1}{l}{POBO}  \\
        $\rho = 1$                                      & $14.8$  &   $15.7$  &   $9.5$    &  $21.8$  \\
        $\rho = 2$                                      & $14.8$  &   $15.9$  &   $8.7$    &  $22.3$  \\
        $\rho = 5$                                      & $14.8$  &   $15.7$  &   $5.1$    &  $21.9$  \\
        $\rho = 10$                                     & $14.8$  &   $16.0$  &   $4.4$    &  $18.8$  \\             
        \midrule 
        \multicolumn{1}{l}{Minimal tightening}          & $0$    &   $12.3$    &  \textemdash &   \textemdash  \\
        \multicolumn{1}{l}{Maximal tightening}          & $41$   &   $19.0$    &  \textemdash &   \textemdash  \\
        \bottomrule
    \end{tabular}
\end{table}

Surprisingly, we find that maximal tightening does not produce the most damaging attack.
Rather, except for CABO and POBO with $\rho = 10$, all other experiments find more damaging attacks, including naive random sampling.
In fact, unconstrained BO can find substantially more damaging attacks compared to both random sampling and maximal tightening, which is particularly promising in demonstrating the viability of using BO for solving this problem given the complete absence of hyperparameter tuning.
Even within the constrained setting, we observe all POBO experiments outperforming a feasible random sampling baseline, although CABO is unable to do so--we surmise that this is because trust region based constrained optimization of the acquisition function can be too restrictive for CABO to sample effective regions of the search space.
For POBO however, our experiments show that it is still possible to find damaging attacks with substantially less overall tightening.
For instance, POBO with $\rho = 2$ finds at attack that deals $97\%$ of the cascade damage from the best unconstrained attack, but does so with $68\%$ less tightening.
That is, even if an adversary was severely constrained in their ability to tighten lines, they would still be able to find an attack that deals roughly the same amount of cascade damage as if they were not constrained.

\subsection{Progression of evaluations} \label{sec:experiments:progression_evaluations}
\begin{figure}[htb]
    \centering
    \captionsetup[subfigure]{font={scriptsize},justification=centering}
    \subfloat[][Random sampling. \label{fig:random_sampling}]{%
        \resizebox{0.24\linewidth}{!}{\includegraphics{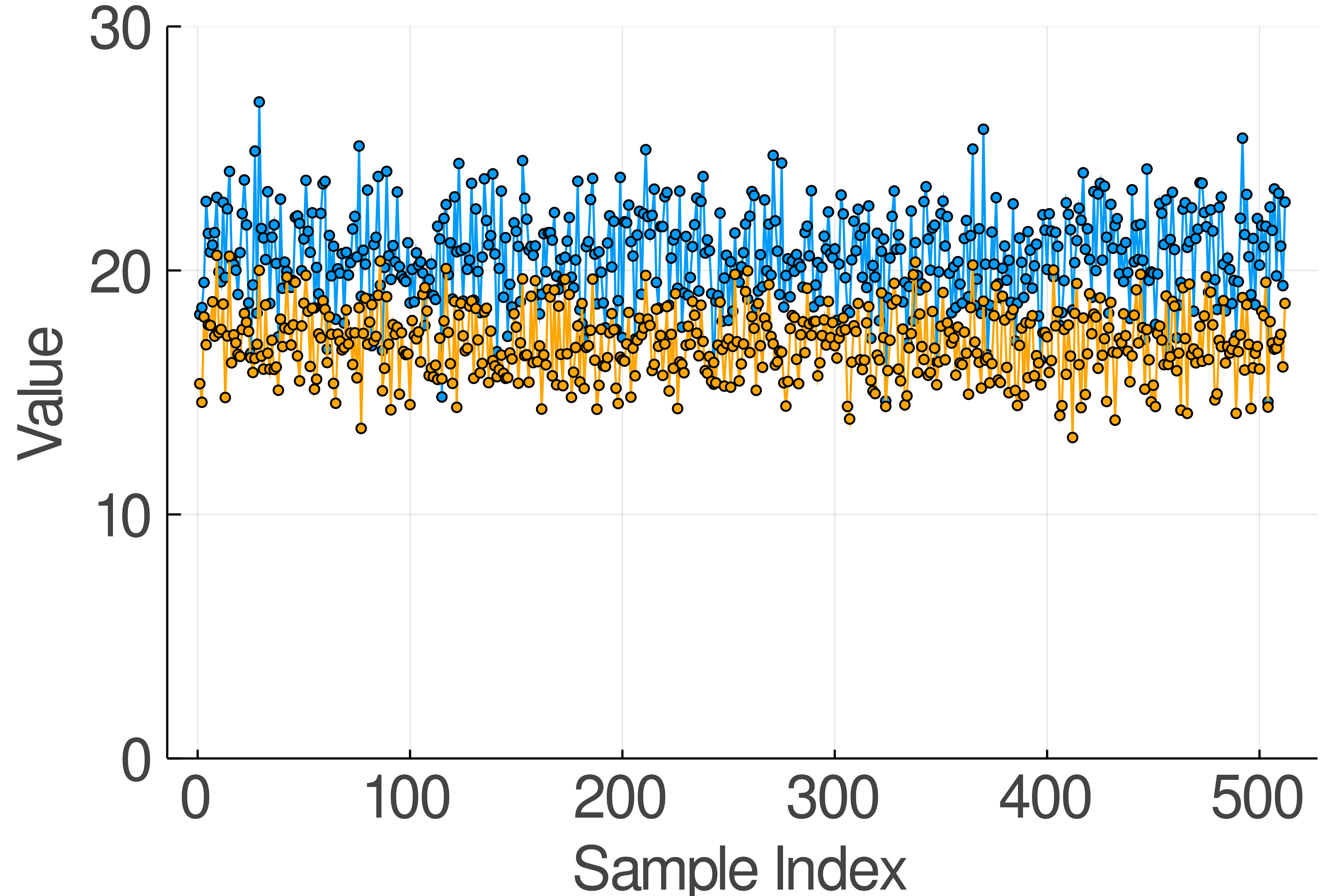}}
    }\hfil
    \subfloat[][Unconstrained BO. \label{fig:unconstrained_BO}]{%
        \resizebox{0.24\linewidth}{!}{\includegraphics{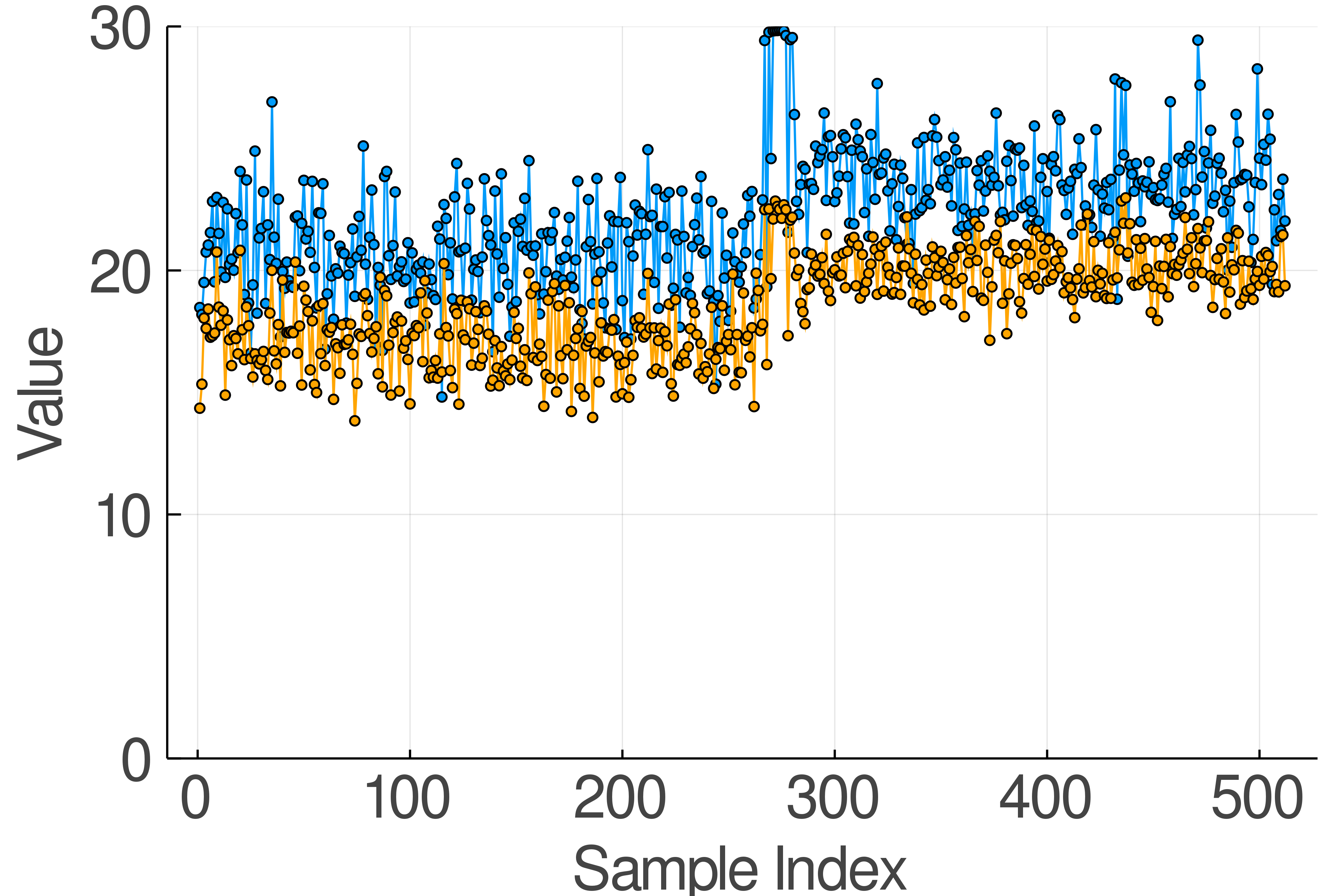}}
    }\hfil
    \subfloat[][Feasible random sampling. \label{fig:feasible_constrained}]{%
        \resizebox{0.24\linewidth}{!}{\includegraphics{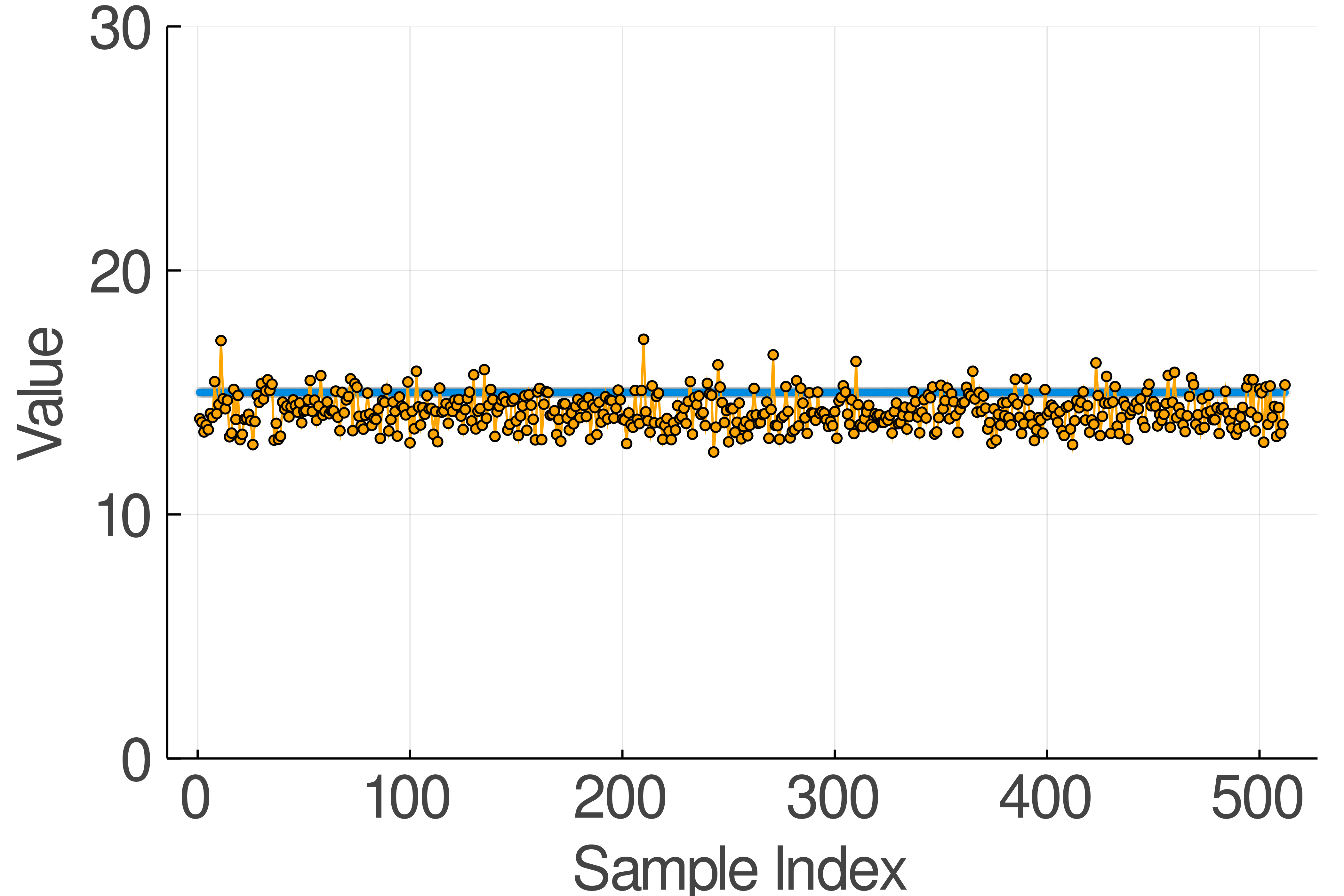}}
    }\hfil
    \subfloat[][CABO. \label{fig:trust}]{%
        \resizebox{0.24\linewidth}{!}{\includegraphics{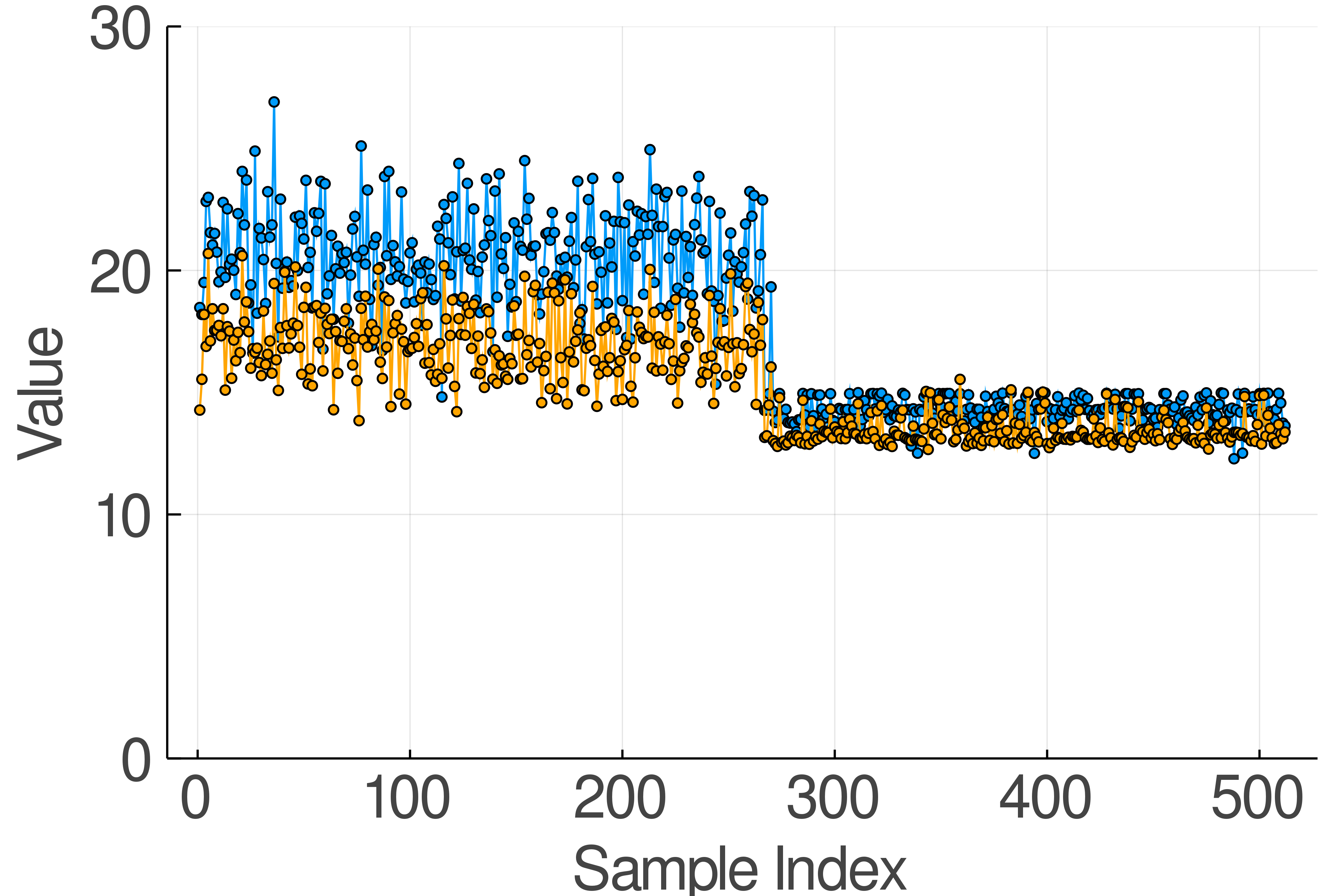}}
    }\hfil

    \subfloat[][POBO with $\rho = 1$. \label{fig:penalized_BO_1}]{%
        \resizebox{0.24\linewidth}{!}{\includegraphics{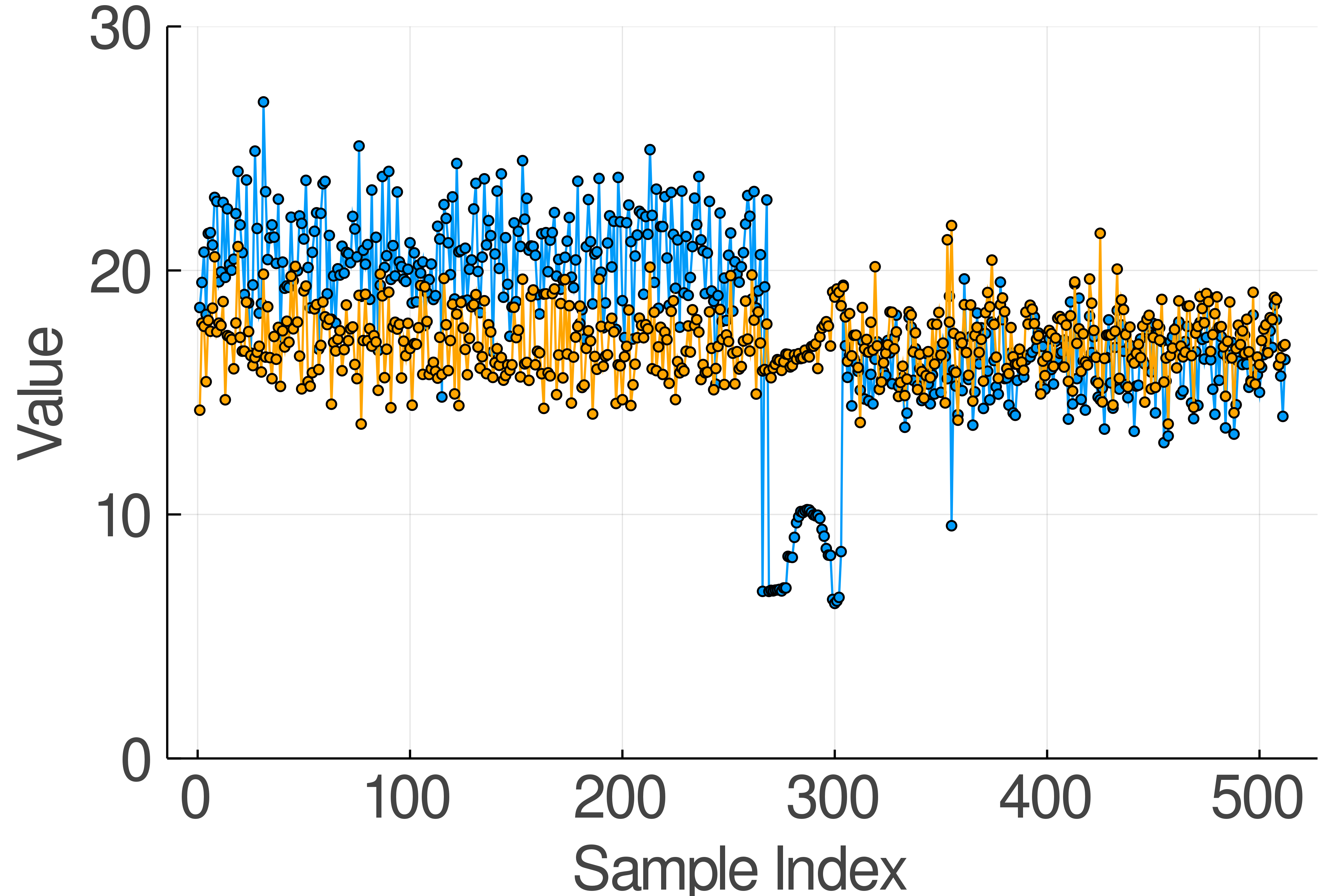}}
    }\hfil
    \subfloat[][POBO with $\rho = 2$. \label{fig:penalized_BO_2}]{%
        \resizebox{0.24\linewidth}{!}{\includegraphics{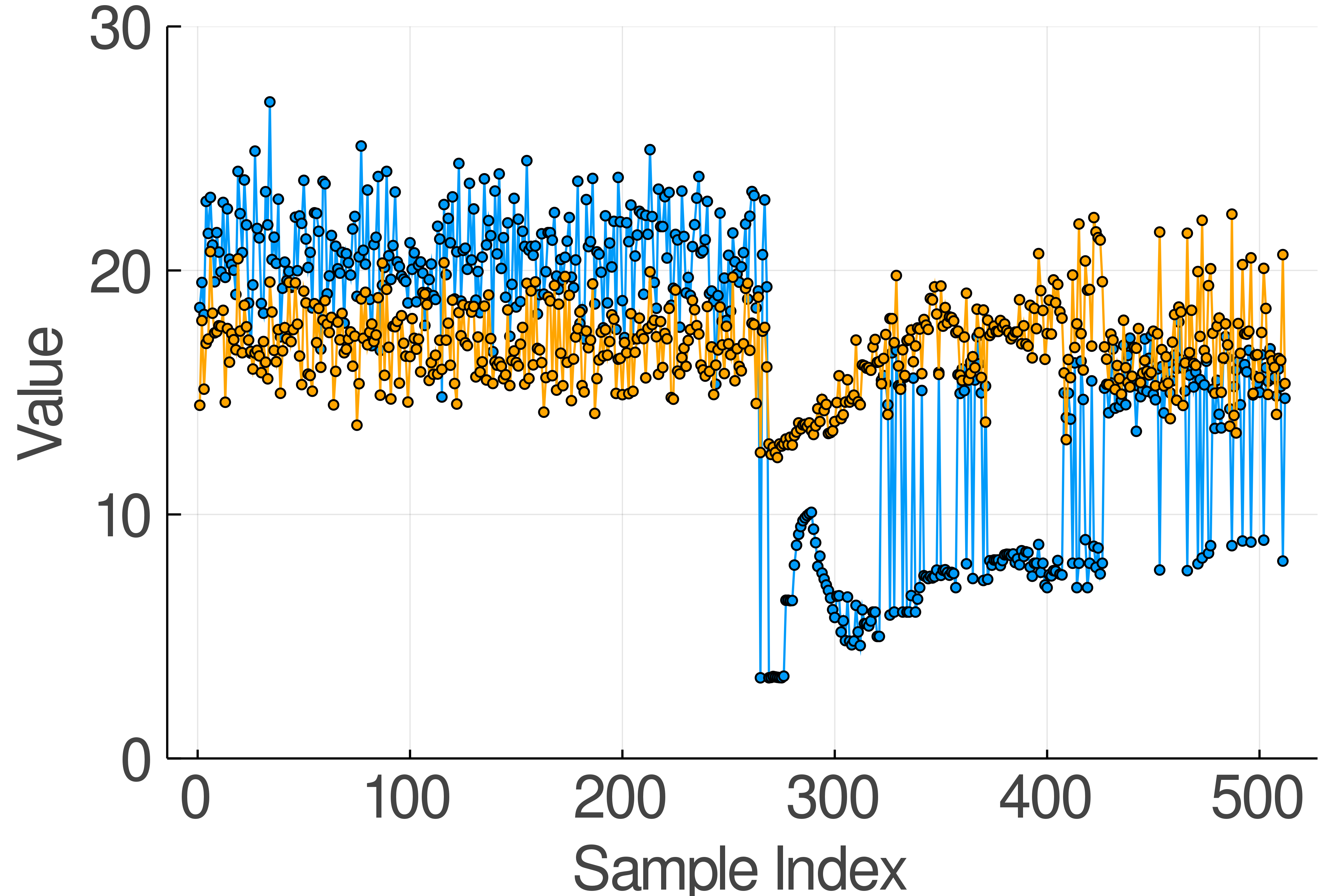}}
    }\hfil
    \subfloat[][POBO with $\rho = 5$. \label{fig:penalized_BO_5}]{%
        \resizebox{0.24\linewidth}{!}{\includegraphics{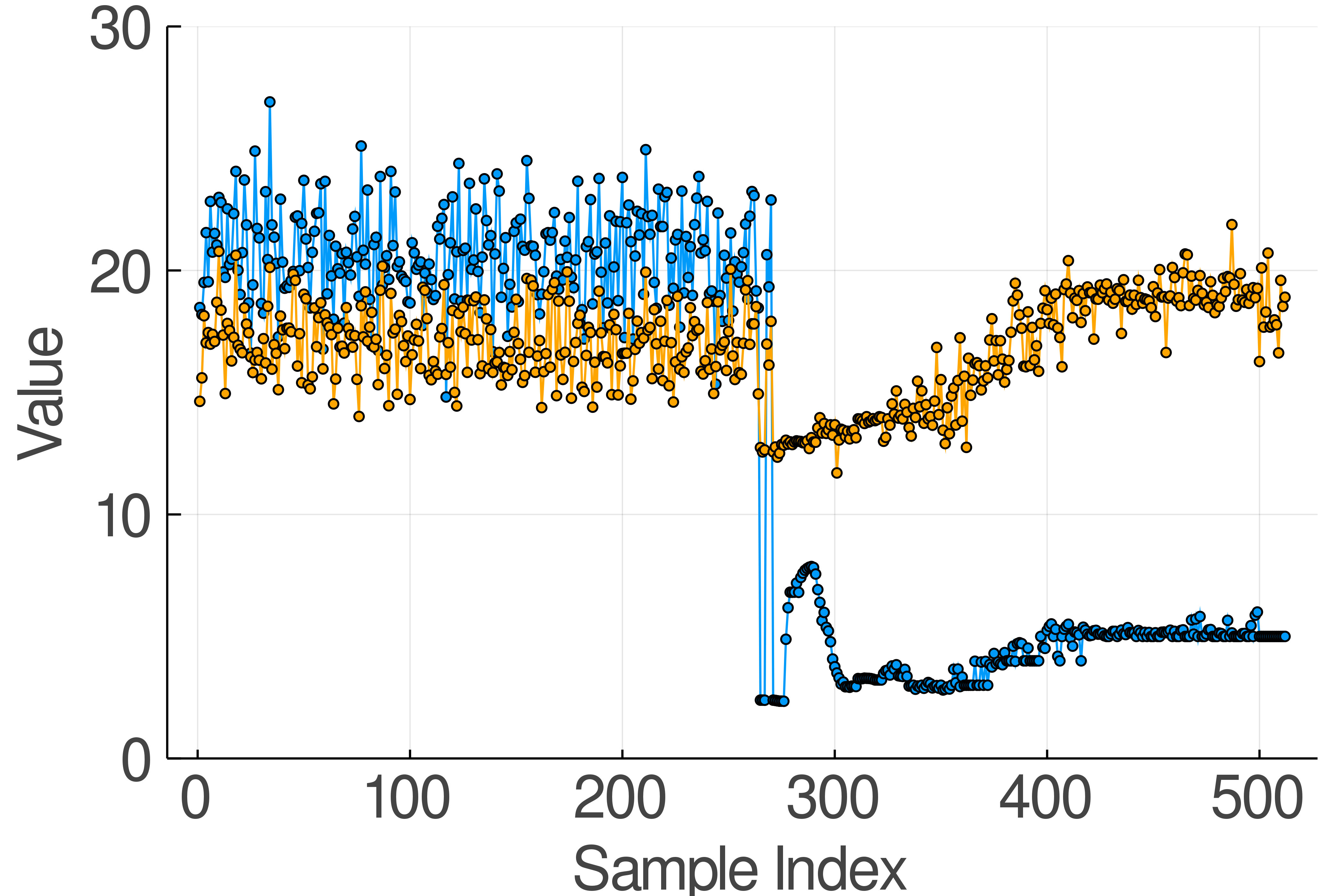}}
    }\hfil
    \subfloat[][POBO with $\rho = 10$. \label{fig:penalized_BO_10}]{%
        \resizebox{0.24\linewidth}{!}{\includegraphics{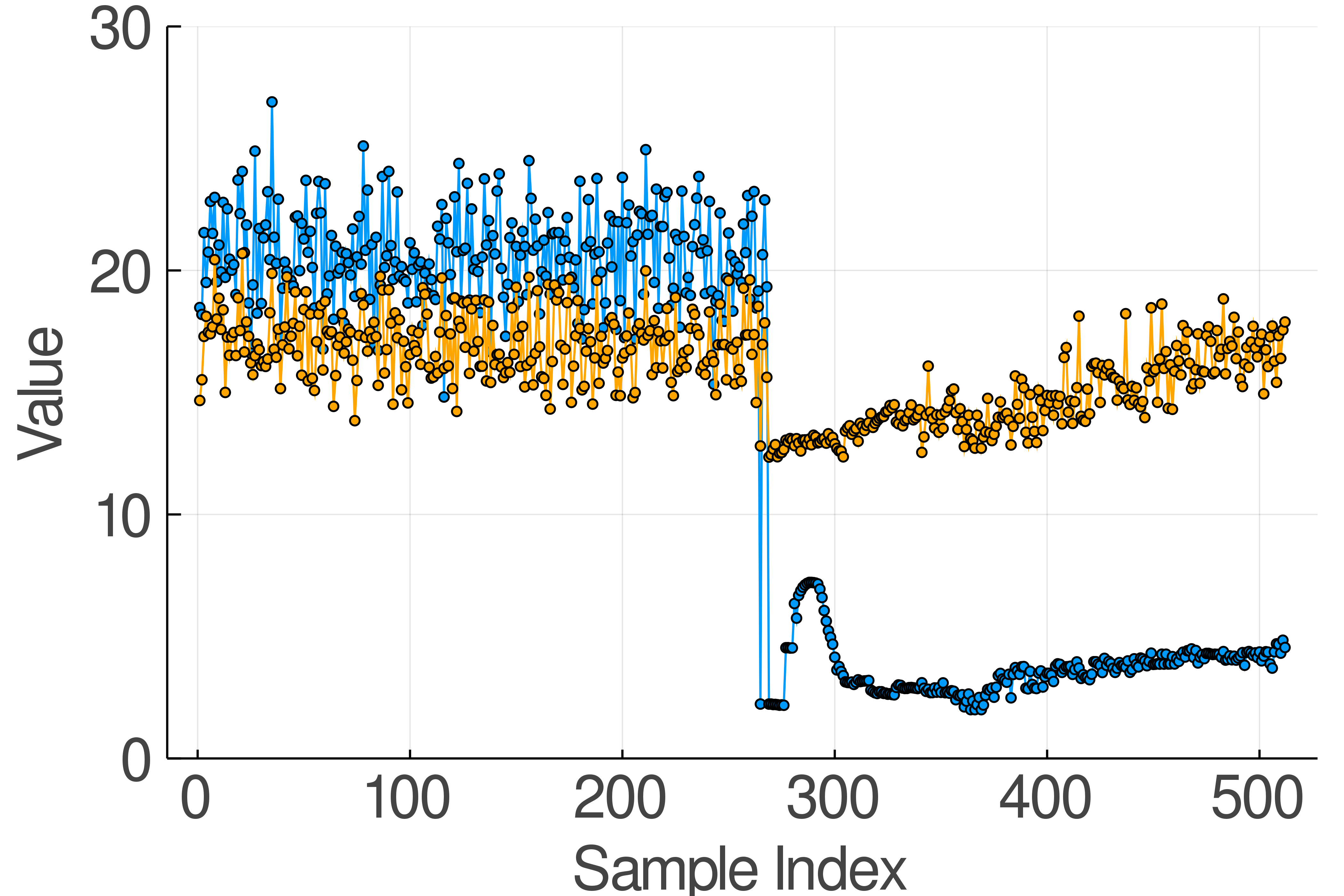}}
    }

    \caption{Progression of objective value $f(x)$ (in orange) and $\|x\|_1$ (in blue) across evaluations.}
    \label{fig:eval_progression}
\end{figure}

Figure~\ref{fig:eval_progression} shows the progression across each evaluation of the objective $f(x)$ (in orange) and configuration norm $\|x\|_1$ (in blue) for the experiments listed in Table~\ref{table:best_empirical_attacks}.
Although there is no constraint in the random sampling or unconstrained BO experiments, we still plot $\|x\|_1$ to compare the amount of tightening present in these sampled configurations with the ones found by the CABO and POBO procedures.

Figures~\ref{fig:random_sampling} and~\ref{fig:feasible_constrained} demonstrate how randomly sampling line limit configurations is an ineffective way for finding potent attacks, with all $512$ evaluations being within a narrow band of objective values.
In contrast, Figure~\ref{fig:unconstrained_BO} shows a noticeable improvement in the objective as soon as the BO routine is activated after the $256$ initial evaluations.
It is noteworthy that this increase in the objective broadly coincides with tighter line limits.

The POBO experiments in Figures~\ref{fig:penalized_BO_1}-\ref{fig:penalized_BO_10} show the importance of tuning $\rho$ to balance the trade-off between maximizing the objective and sampling feasible configurations that do not incur the cost of a penalty, with larger values of $\rho$ encouraging feasibility, albeit generally at the cost of lower objective values.
However, some of these experiments that use larger penalties are still able to find very damaging attacks, as seen in Figure~\ref{fig:penalized_BO_5}.
Comparing this with Figure~\ref{fig:penalized_BO_1} demonstrates that tuning $\rho$ is important: setting $\rho = 5$ finds an attack that marginally outperforms the one found when setting $\rho = 1$, even though it does $46\%$ less tightening, while the best attack found by using $\rho = 1$ outperforms the one found by using $\rho = 10$ by $16\%$, but does so with more than twice the amount of tightening. 

\subsection{Attack sparsity and evaluation time}
\begin{figure}[htb]
    \centering
    \captionsetup[subfigure]{font={scriptsize},justification=centering}
    \subfloat[][Feasible random sampling (red) vs  Unconstrained BO (grey). \label{fig:alpha_dist_unconstrained}]{%
        \resizebox{0.24\linewidth}{!}{\includegraphics{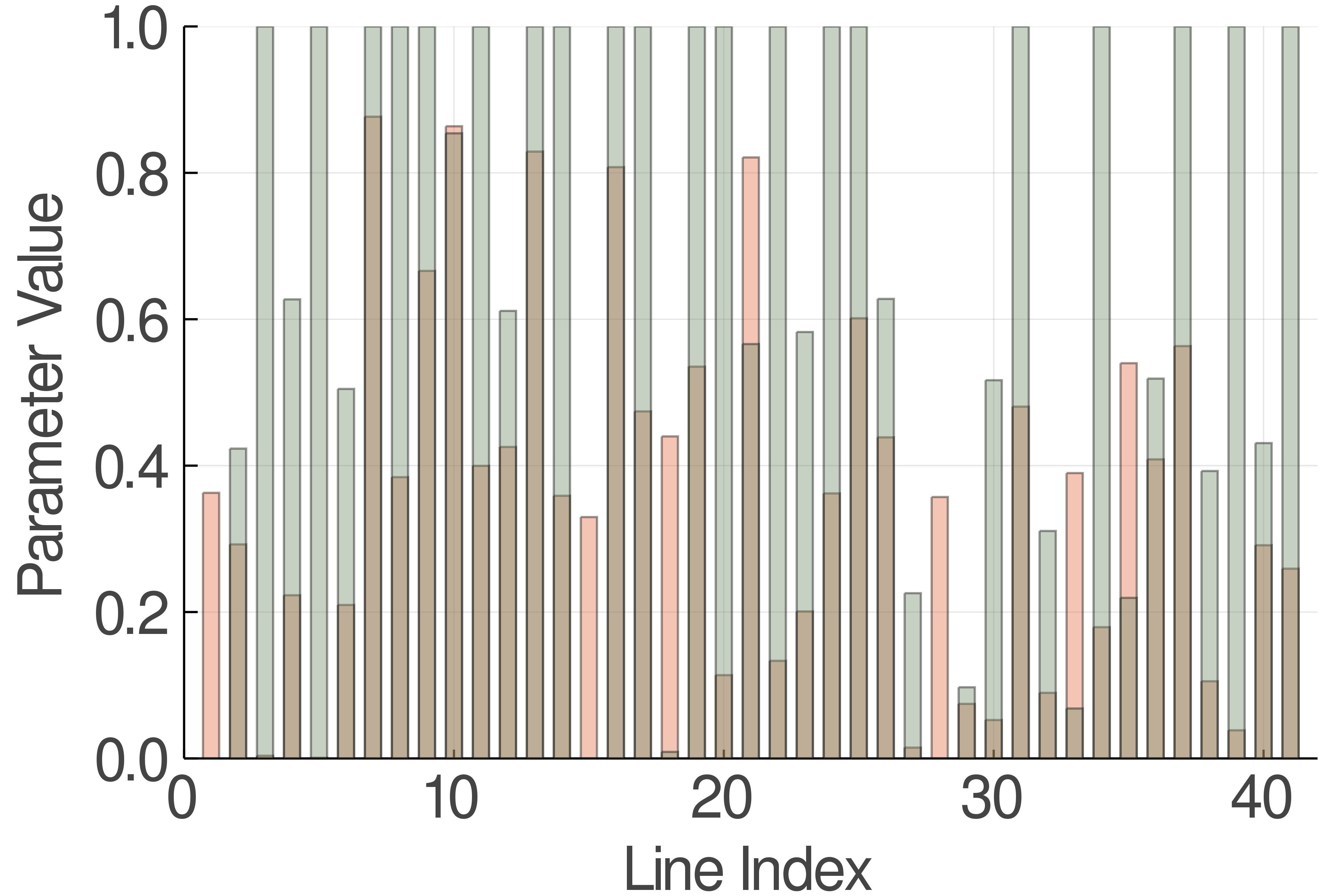}}
    }\hfil
    \subfloat[][Feasible random sampling (red) vs CABO (green). \label{fig:alpha_dist_trust}]{%
    \resizebox{0.24\linewidth}{!}{\includegraphics{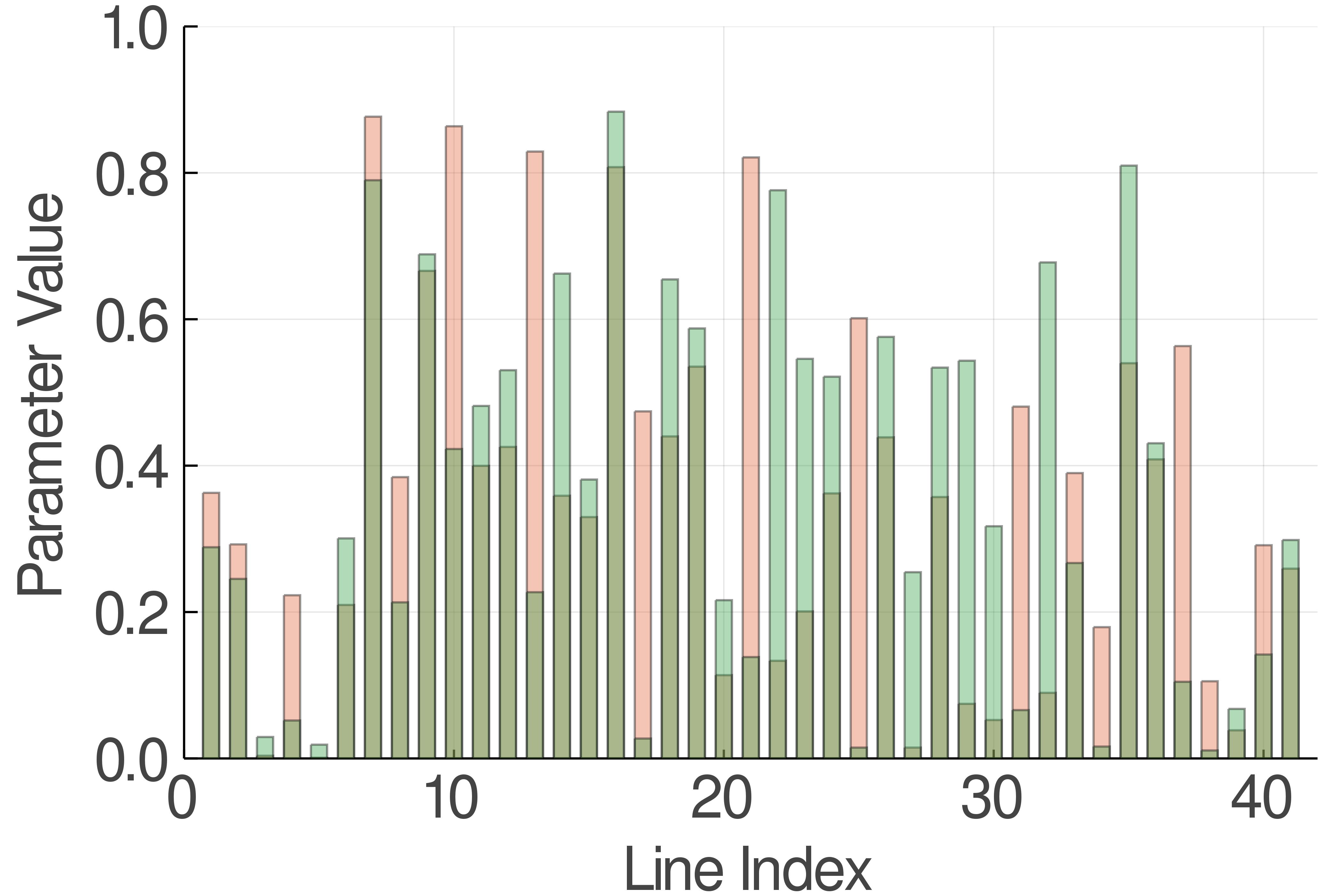}}
    }\hfil
    \subfloat[][Feasible random sampling (red) vs POBO with $\rho = 1$ (tan). \label{fig:alpha_dist_pen_1}]{%
        \resizebox{0.24\linewidth}{!}{\includegraphics{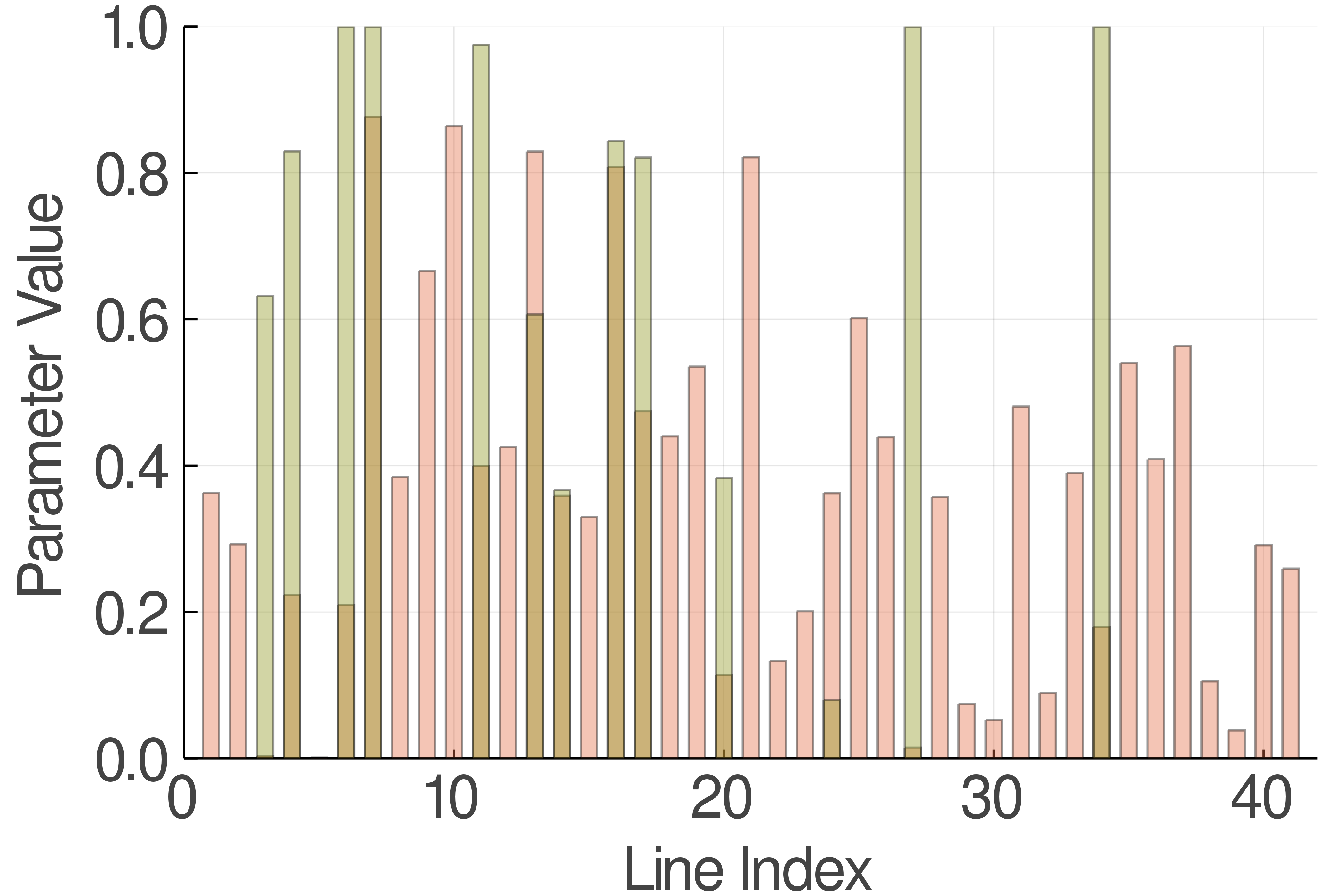}}
    }\hfil
    \subfloat[][Feasible random sampling (red) vs POBO with $\rho = 2$ (purple). \label{fig:alpha_dist_pen_2}]{%
        \resizebox{0.24\linewidth}{!}{\includegraphics{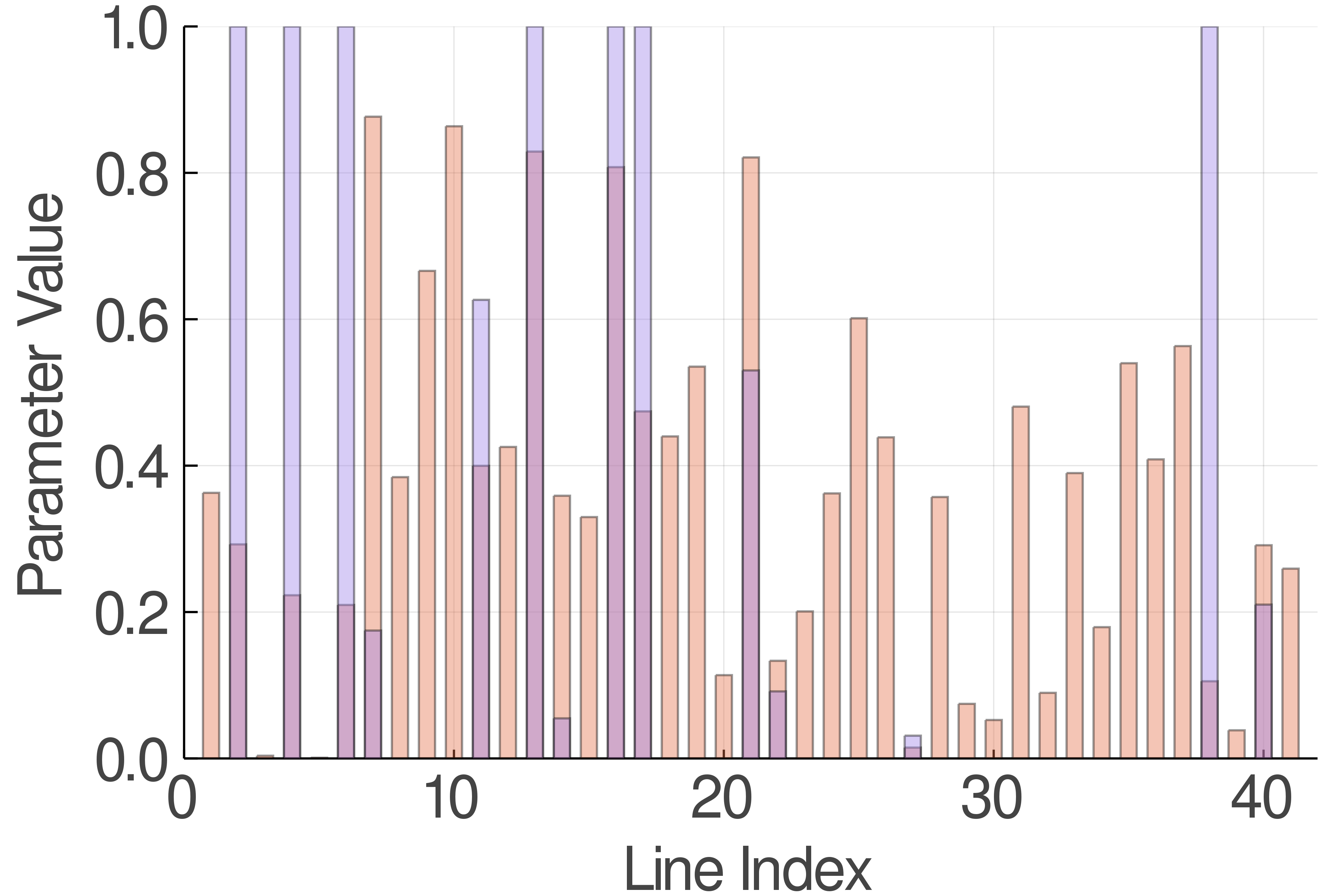}}
    }\hfil

    \subfloat[][Feasible random sampling (red) vs POBO with $\rho = 5$ (teal). \label{fig:alpha_dist_pen_5}]{%
        \resizebox{0.24\linewidth}{!}{\includegraphics{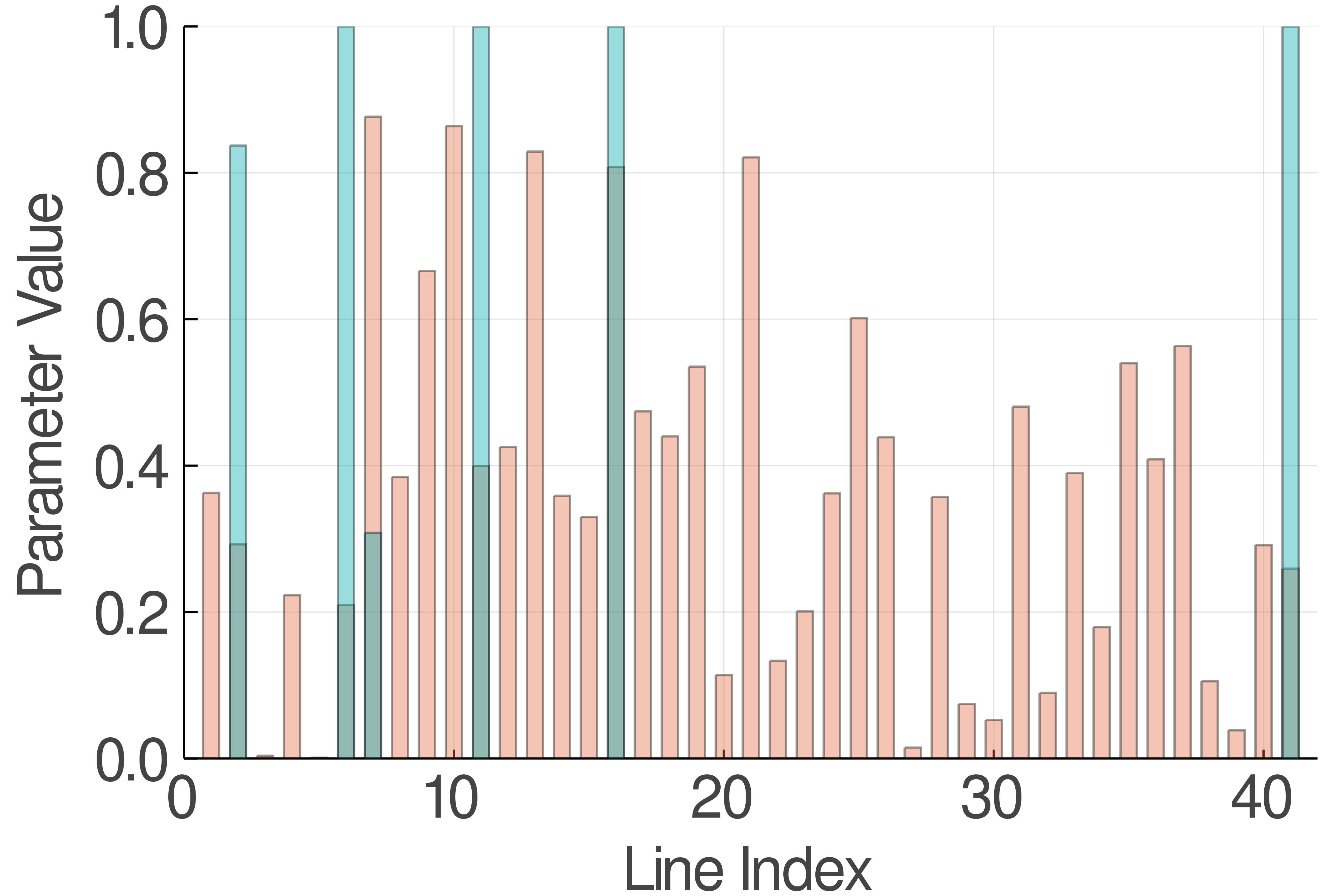}}
    }\hfil
    \subfloat[][Feasible random sampling (red) vs POBO with $\rho = 10$ (salmon). \label{fig:alpha_dist_pen_10}]{%
        \resizebox{0.24\linewidth}{!}{\includegraphics{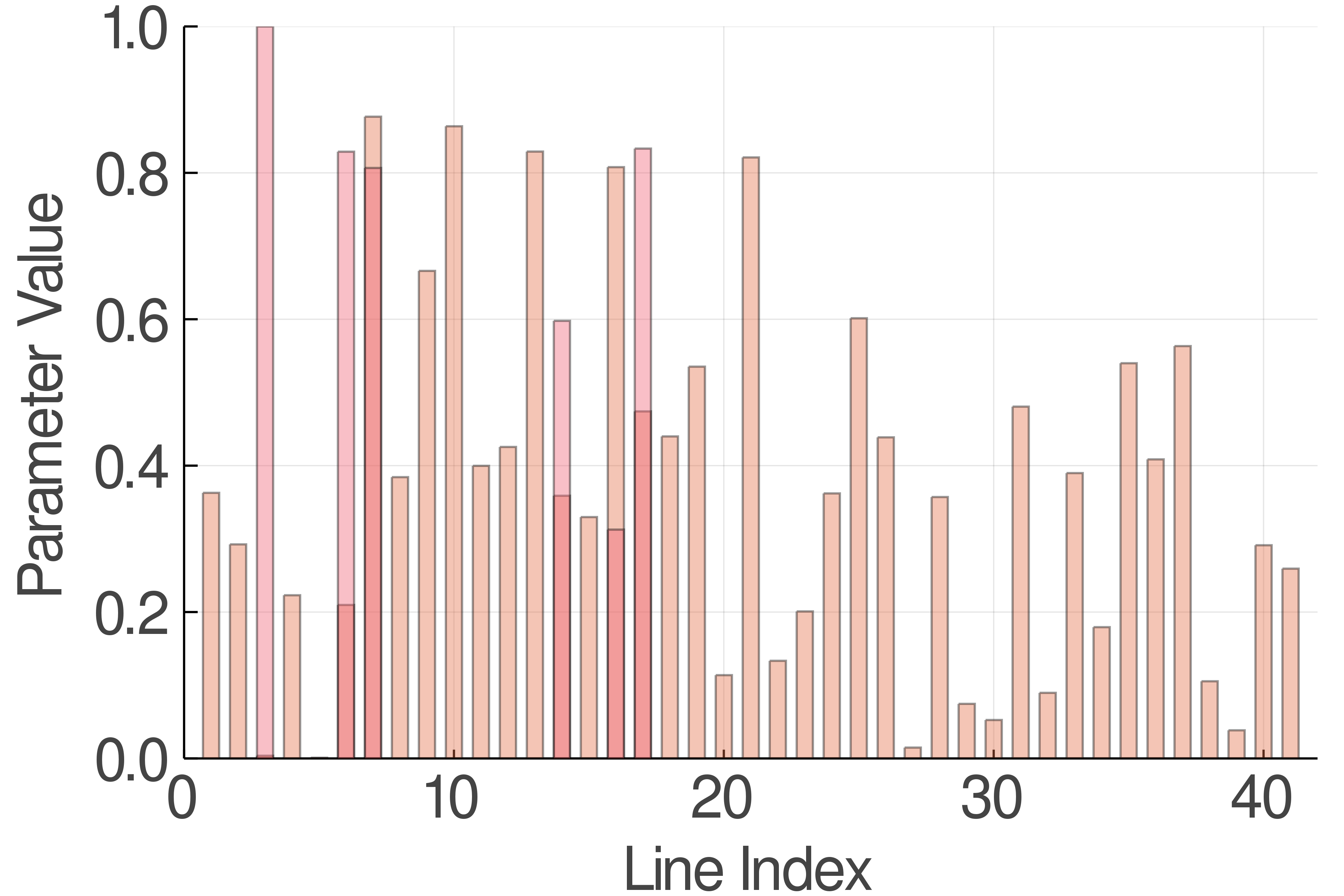}}
    }\hfil
    \subfloat[][Number of non-zero parameters. \label{fig:nnz}]{%
        \resizebox{0.24\linewidth}{!}{\includegraphics{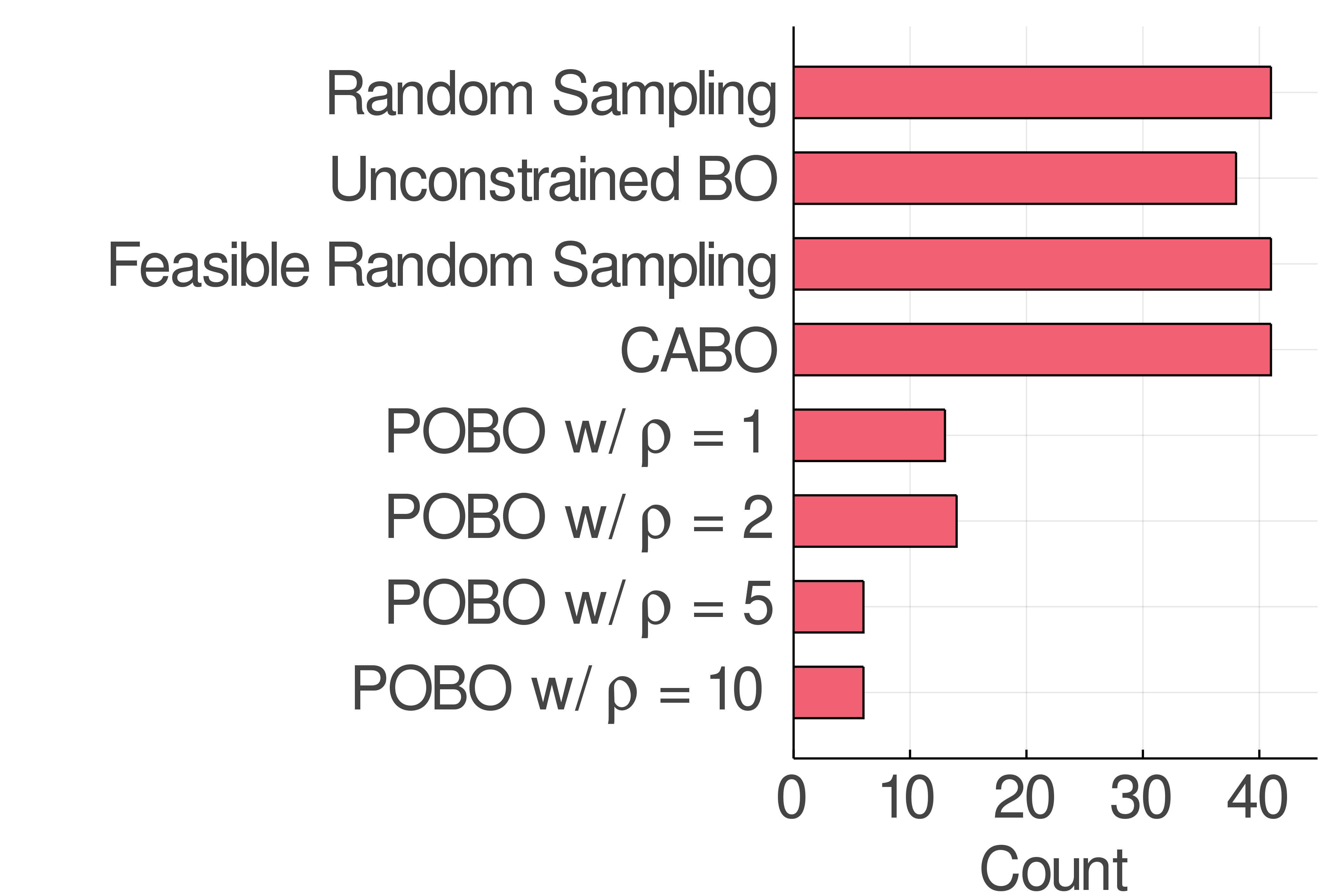}}
    }\hfil
    \subfloat[][Average time (secs) of one BO evaluation. \label{fig:BO_eval_times}]{%
        \resizebox{0.24\linewidth}{!}{\includegraphics{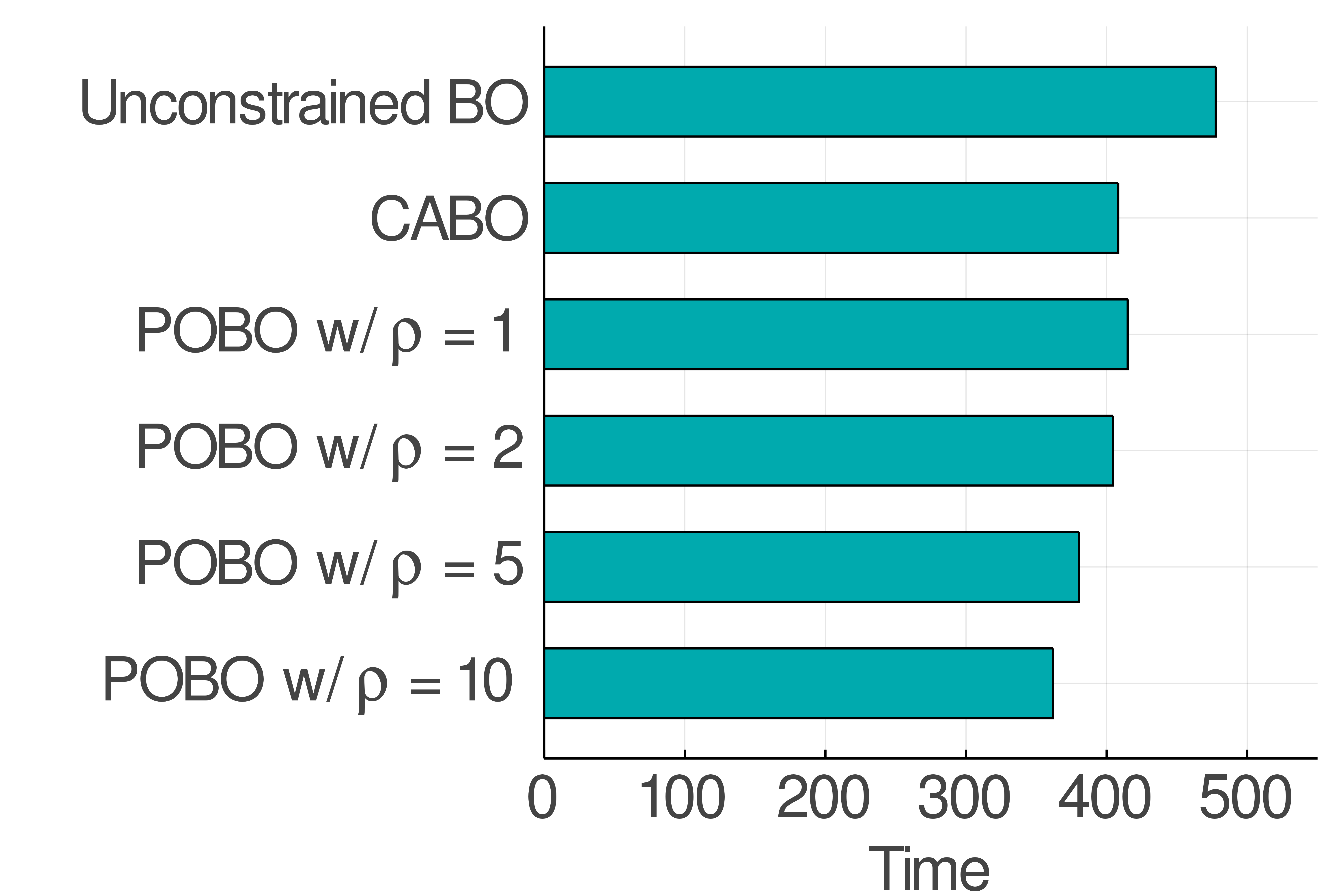}}
    }

    \caption{Distribution of line limit tightening and the evaluation time for the best attack of each experiment.}
\end{figure}

Figures~\ref{fig:alpha_dist_unconstrained}-\ref{fig:alpha_dist_pen_10} show the distribution of tightening across the 41 transmission lines for the best attacks found in the BO experiments. 
To facilitate comparison, the feasible random sampling results (in red) are used as a reference distribution.
We also present the number of non-zero tightening parameters from these attacks in Figure~\ref{fig:nnz}.

We find that the best CABO attack does not induce any additional sparsity over the unconstrained attack---in fact, the unconstrained BO attack has 38 non-zero tightening parameters, compared to the 41 of the CABO attack.
In contrast, Figures~\ref{fig:alpha_dist_pen_1}-\ref{fig:alpha_dist_pen_10} demonstrate how larger values of $\rho$ tend to increase sparsity.
We are however unable to find consistency in which lines are tightened as $\rho$ is increased, nor do we observe correlation between which lines are tightened and the amount of resulting cascade damage, suggesting that multiple, yet unrelated, potent attack vectors may exist for a given network.

We present the average walltime to perform one BO loop in these experiments, which were run using 12 parallel processes, in Figure~\ref{fig:BO_eval_times} and observe that larger values of $\rho$ in the POBO experiments broadly correspond to steady reductions in the average time it takes to make one evaluation.
This is because the set of simulated cascades in each BO loop tends to be shorter on average as $\rho$ increases, as shown in Figure~\ref{fig:eval_progression}, which we expect to be the major contributor to evaluation time since longer cascades typically take longer to simulate.
However, this does not preclude POBO from finding effective attacks as $\rho$ increases, as shown in Table~\ref{table:best_empirical_attacks}.

\subsection{Insights into line failure} \label{sec:experiments:line_failures}
We try to provide some insight into why the best attack found by POBO tends to inflict more cascade damage than minimal and maximal tightening.

\begin{figure}[htb]
    \centering
    \subfloat[][Minimal tightening. \label{fig:min_line_failure_hist}]{%
        \resizebox{0.30\linewidth}{!}{\includegraphics{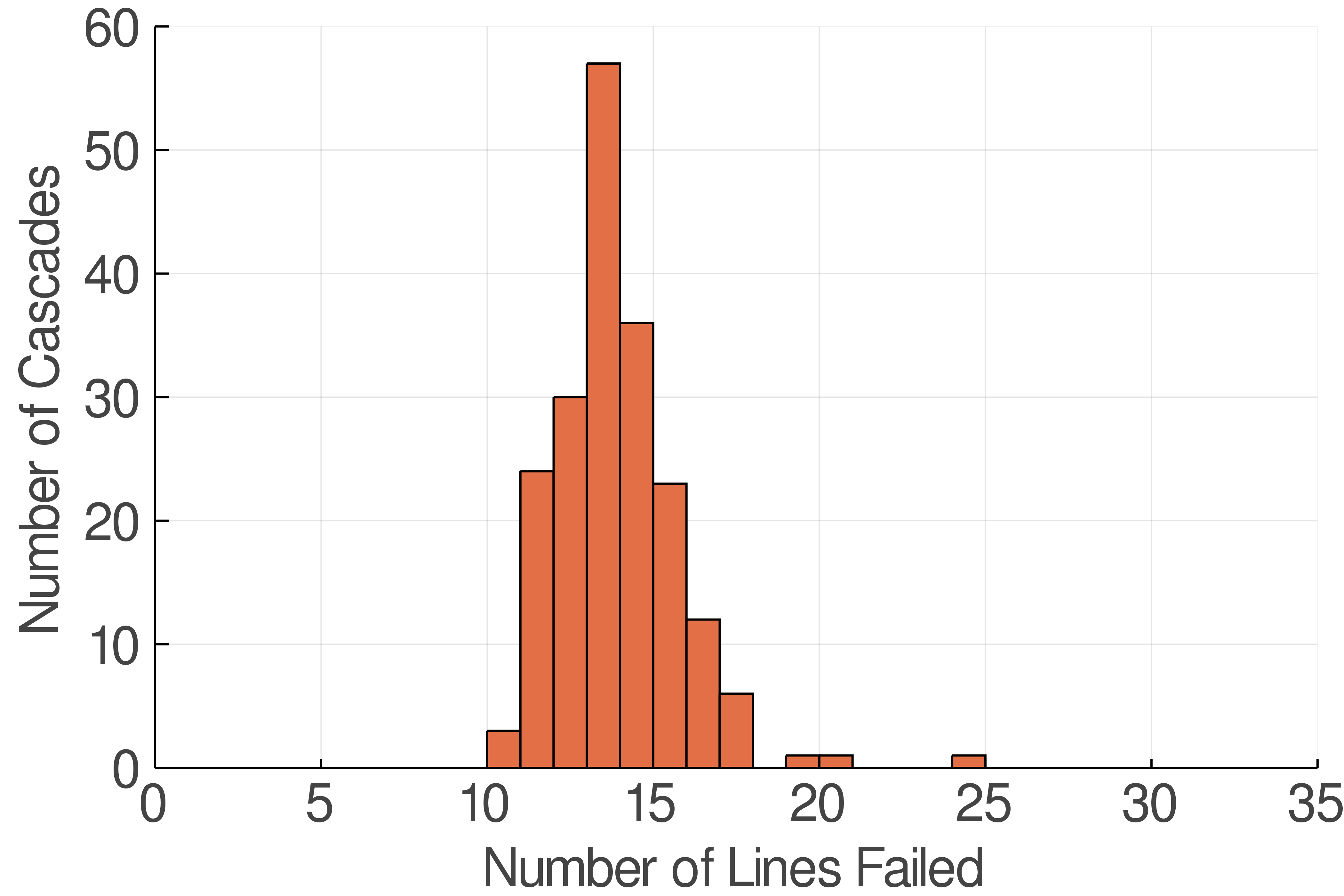}}
    }\hfil
    \subfloat[][Best POBO attack with $\rho = 2$. \label{fig:pen_2_line_failure_hist}]{%
        \resizebox{0.30\linewidth}{!}{\includegraphics{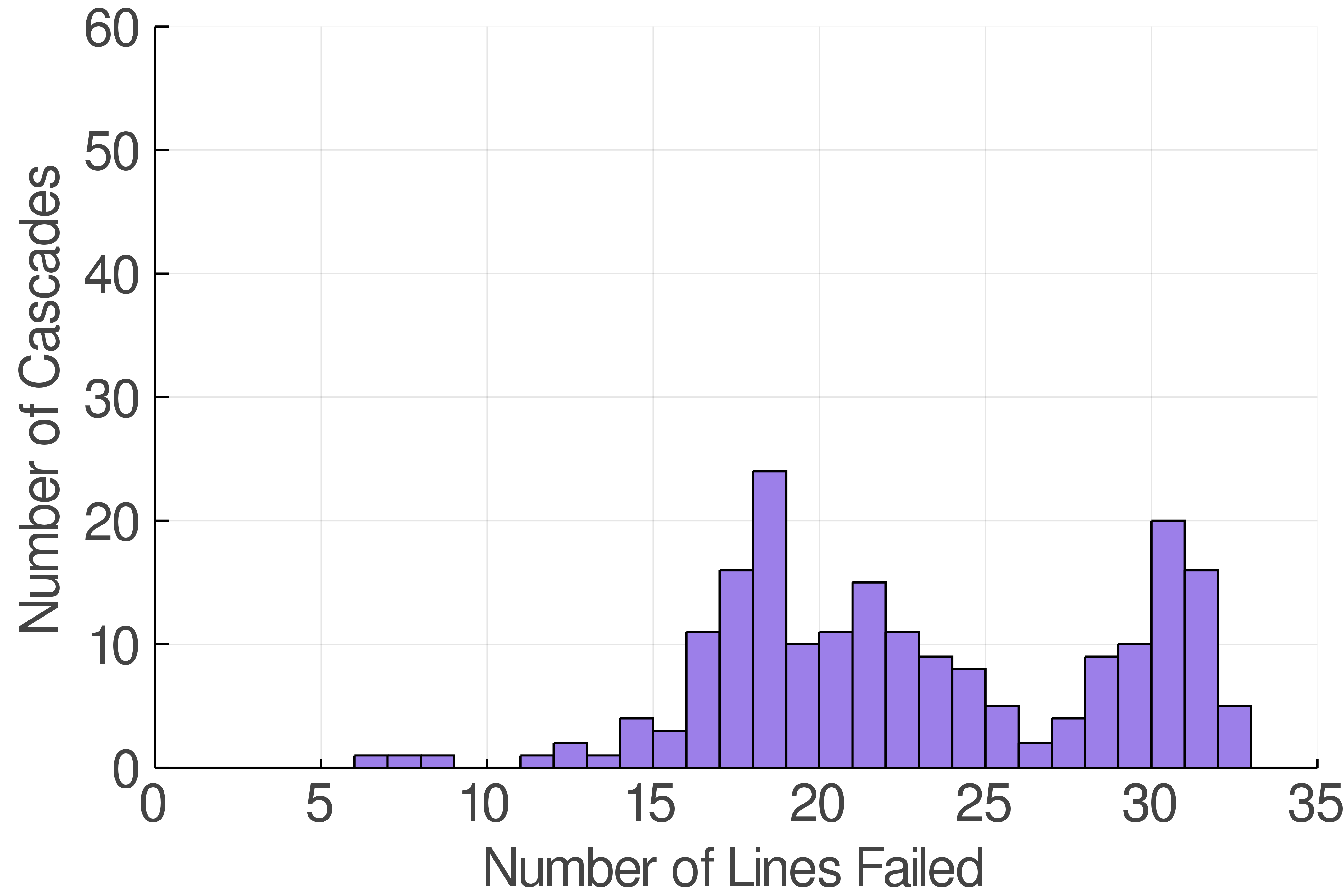}}
    }\hfil
    \subfloat[][Maximal tightening. \label{fig:max_line_failure_hist}]{%
    \resizebox{0.30\linewidth}{!}{\includegraphics{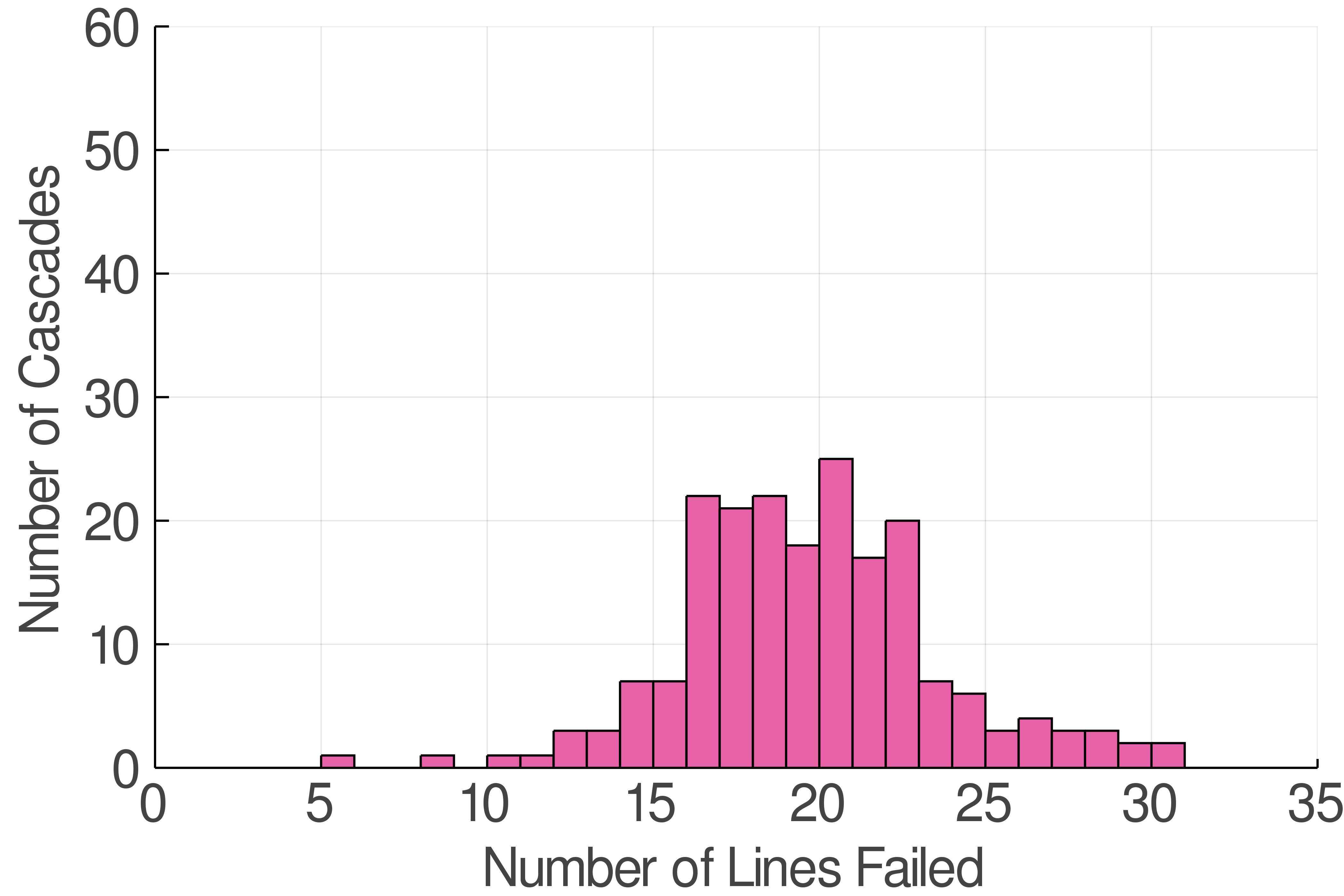}}
    }\hfil
    \caption{Histogram of the number of line failures in a cascade.}
    \label{fig:line_failure_hist}
\end{figure}

First, we plot the histogram of line failures for a set of approximately $200$ cascades (i.e., the number of cascade simulations we used in one BO evaluation for our experiments), comparing the best POBO attack (found using $\rho = 2$) against minimal and maximal tightening.
This is shown in Figure~\ref{fig:line_failure_hist}, where we observe that while all three attacks can produce relatively short cascades, the histogram of the POBO attack is bimodal and is more likely to induce longer cascades.

\begin{figure}[tb]
    \centering
    \subfloat[][Minimal tightening. \label{fig:min_line_failure_analysis}]{%
        \resizebox{0.30\linewidth}{!}{\includegraphics{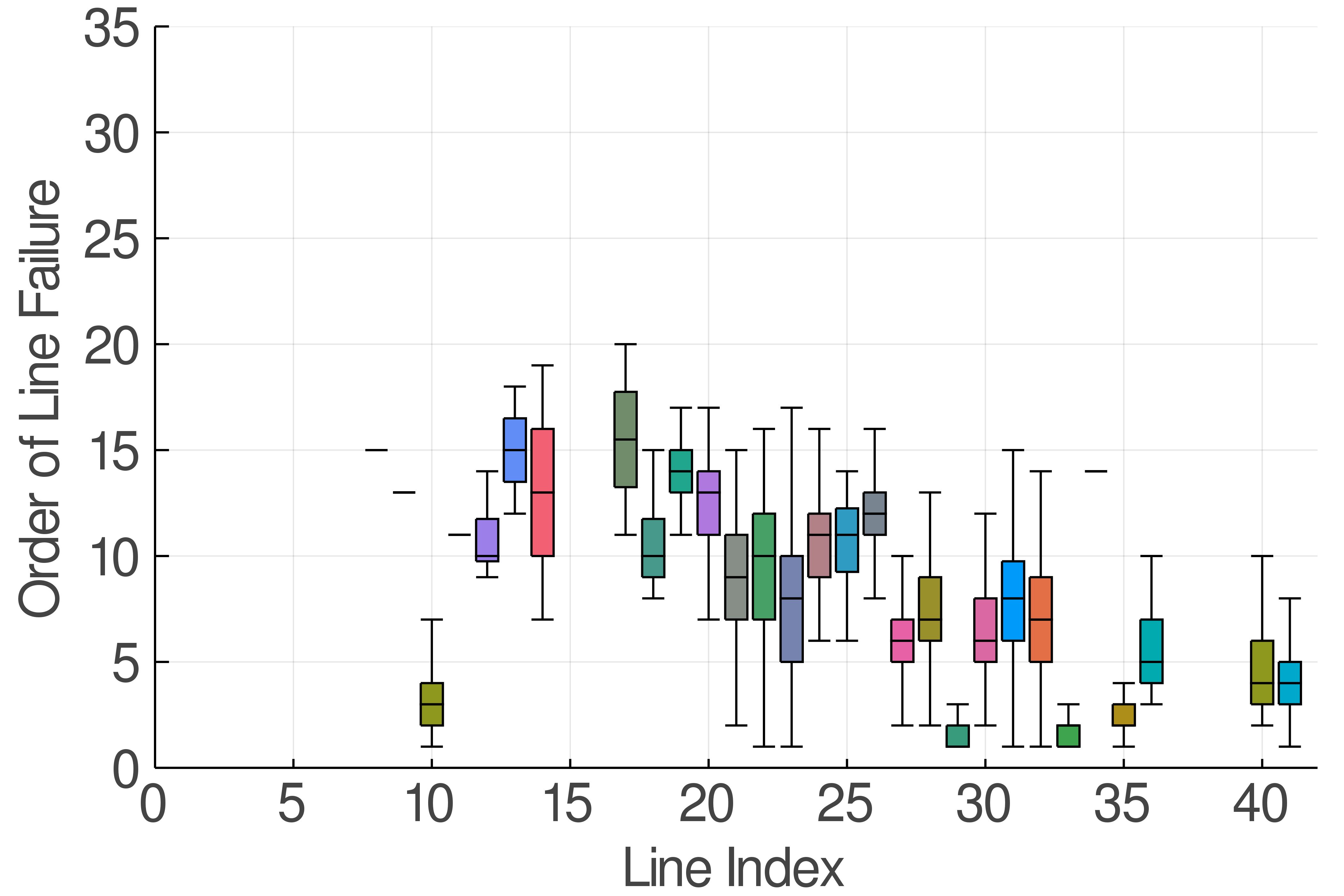}}
    }\hfil
    \subfloat[][Best POBO attack with $\rho = 2$. \label{fig:pen_2_line_failure_analysis}]{%
        \resizebox{0.30\linewidth}{!}{\includegraphics{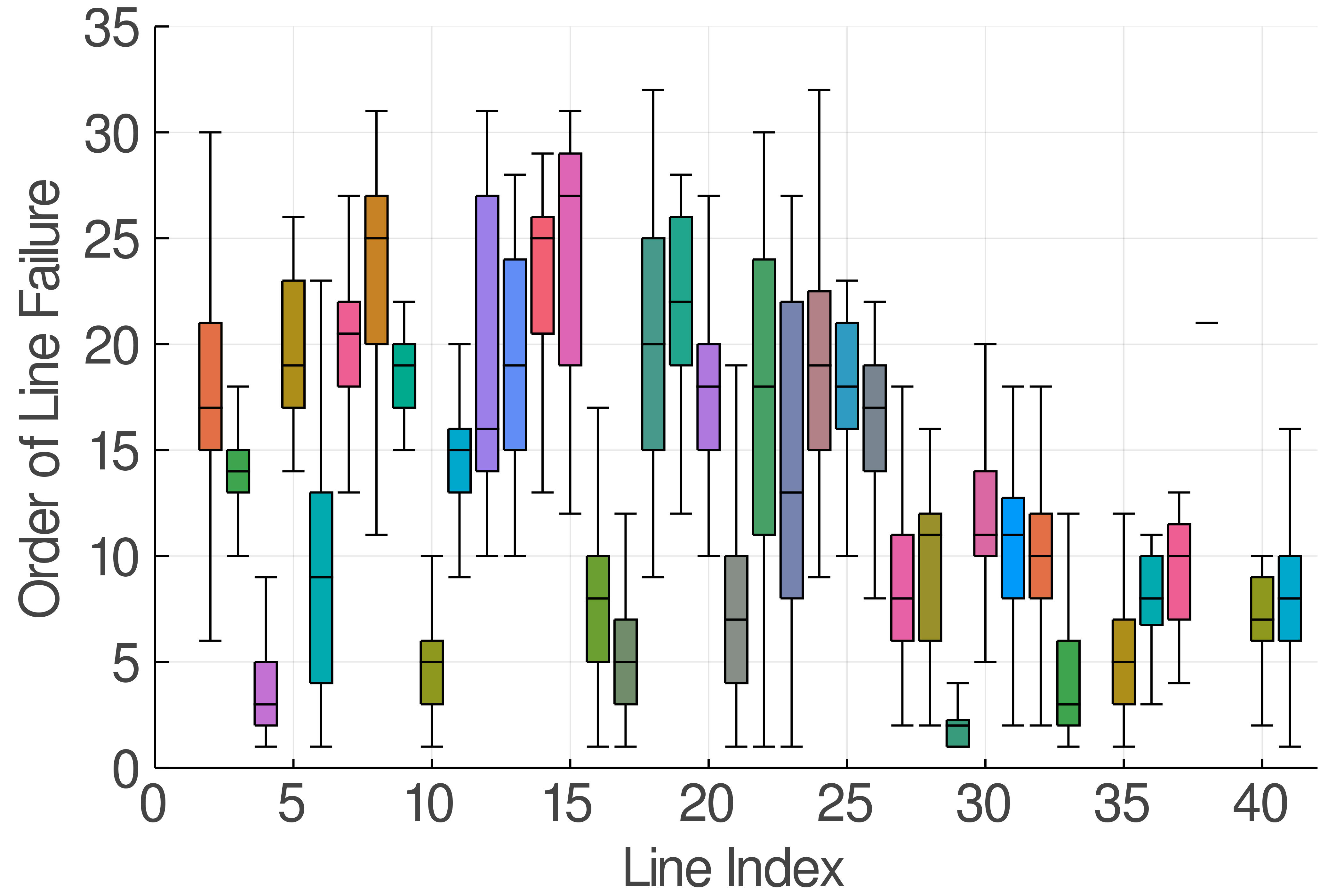}}
    }\hfil
    \subfloat[][Maximal tightening. \label{fig:max_line_failure_analysis}]{%
    \resizebox{0.30\linewidth}{!}{\includegraphics{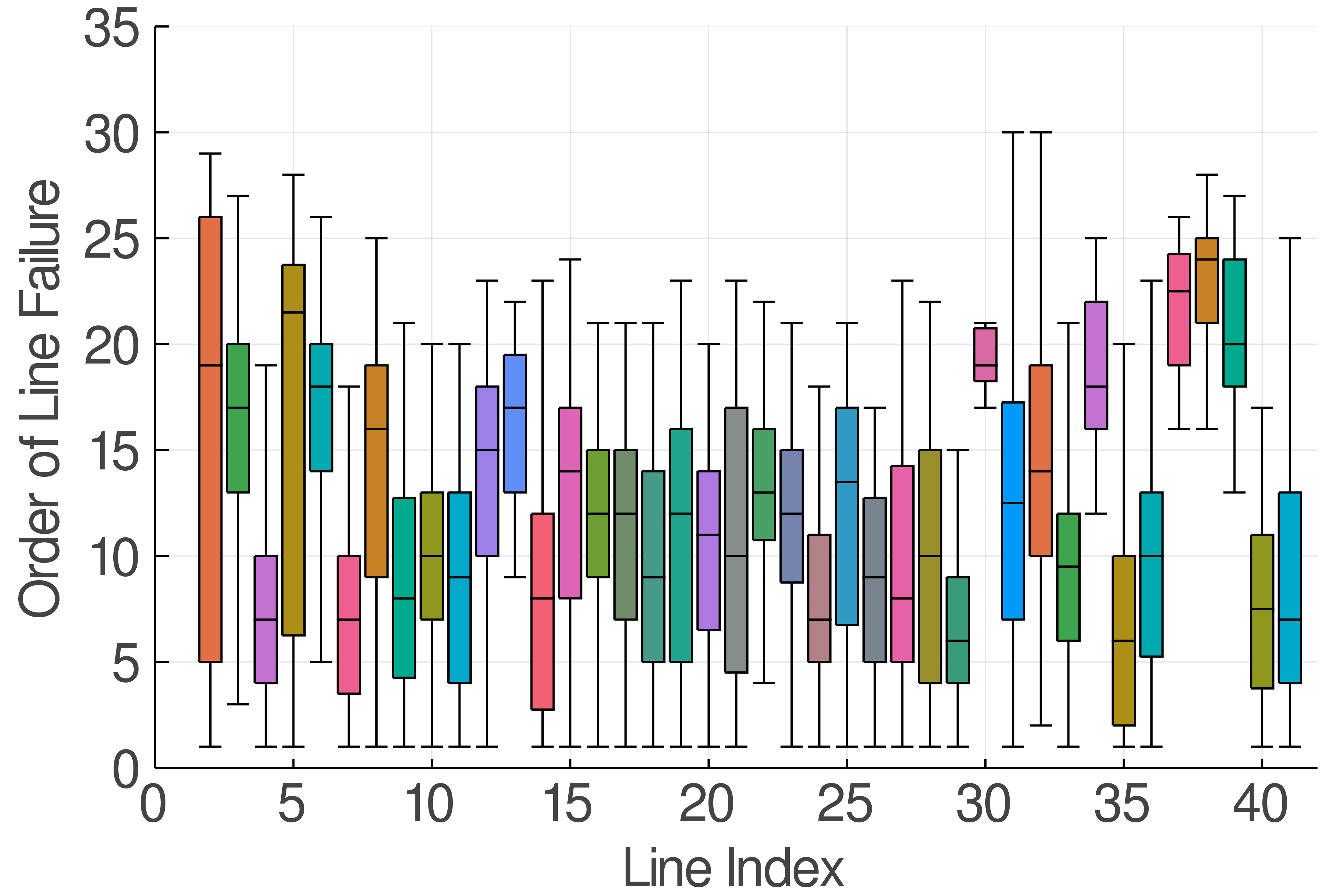}}
    }
    \caption{Order of line failure within the cascade across each line of the network.}
    \label{fig:line_failure_analysis}
\end{figure}

We then plot the order in which each line of the network fails within the cascade for each case. 
Figures~\ref{fig:min_line_failure_analysis}-\ref{fig:max_line_failure_analysis} show how the order of each line failure tends to be more concentrated along each line index when there is less overall tightening.
That is, more tightening tends to produce line failure sequences that are generally less predictable.
So while minimal tightening tends to fail lines in a predictable order that produces short cascades, and the less predictable failure order of maximal tightening will tend to produce longer cascades, it seems the tightening applied by the BO attack tends to fail lines in an order that can occasionally produce even longer cascades than the ones induced by maximal tightening, even though there is less overall tightening in this attack.
We surmise that there is some interplay between the amount of tightening and the resulting order in which lines fail that can lead to very damaging attacks, even if only occasionally. 

\section{Future Work} \label{sec:future_work}
We envision a number of research directions that extend our ideas using more recent advances in BO methodology.

Firstly, reducing the dimensionality of the problem would enable our approach to scale to larger networks, which is particularly challenging using standard BO methods, since the dimensionality increases with the number of lines in the network and BO is susceptible to the curse of dimensionality.
One line of research in high-dimensional BO centers on performing BO in a low-dimensional subspace before transforming the points back into the original space using a specially constructed embedding~\citep{wang2016rembo,nayebi2019hesbo}.
Alternatively, one can directly induce sparsity in the surrogate model based on the assumption that optimizing the objective only depends on a small subset of the parameters comprising the search space~\citep{eriksson2021saasbo}.
For the relatively small network in our experiments, we have seen that tightening attacks can still be very effective even when a small subset of lines are tightened, suggesting that the effective dimensionality of the problem may be substantially lower than the actual number of lines in the network.
If this can also be shown in larger networks, these types of methods could be particularly promising in extending our approach to networks with hundreds of lines.
Another avenue of high-dimensional BO focuses on dividing the global optimization problem into localized subproblems using trust regions, in order to encourage better exploitation of the more promising regions of the original search space~\citep{eriksson2019turbo,eriksson2021scbo}, which could be an alternative approach if the effective dimensionality of these problems are still too high when applied to larger networks.

Secondly, the main contributor to the computational cost in our experiments stems from the large number of cascade simulations required to achieve a good estimate of cascade severity for a given set of tightening parameters, since the underlying KMC simulator is stochastic in nature.
This will be increasingly problematic in larger networks, since the cost of each simulation also scales with the number of lines.
To this end, multi-fidelity BO methods may offer more cost-effective ways to solve the optimization problem, as the number of simulations used to evaluate the objective can be incorporated as a fidelity parameter in these settings, thereby enabling cheaper, lower-fidelity evaluations to be exploited and providing an alternative to always having to rely on high-fidelity evaluations~\citep{song2019mfbo,wu2020mfkg}. 

Finally, the desire for sparsity could be directly encoded in a multi-objective BO problem, which would enable practitioners to systematically and efficiently analyze the trade-offs between sparsity and performance, especially in the high-dimensional setting when such methods are also combined with dimensionality reduction techniques~\cite{liu2023sparseBO,daulton2022morbo}.

\section{Conclusion} \label{sec:conclusion}
We believe our results demonstrate the potential for BO to be a useful tool for practitioners in understanding how cascading can manifest, given the innate complexities of cascade simulation within power systems analysis.
Our paper proposes a framework for casting the optimization problems found in the existing vulnerability analysis literature in to the language of BO, where the black-box quality of cascading naturally assumes the role of the objective in unconstrained and constrained BO problems.
Moreover, practitioners are not bound by the modeling choices made in this paper, and each stage could readily be substituted with other alternatives more suitable to the specific problem at hand.

For our chosen problem, we adopted the role of an adversary intent on maximizing cascade damage via remote tampering of transmission lines.
Our experiments show a number of unexpected results, which we argue are unlikely to be discovered using handcrafted approaches given the dimensionality of the search space combined with the complexities of cascade simulation.
For instance, we show that other line limit configurations exist in both the unconstrained and constrained settings that lead to substantially more cascading than the maximal tightening attack, which we would have expected to be the most effective attack.
In fact, we are able to find configurations that have very little tightening while also dealing close to the same amoount of cascade damage as the best empirical solution found across all our experiments, demonstrating that tremendous damage can still be exacted on these networks even when the range of attack vectors is severely restricted.
Some of these attacks also benefit further from sparsity, which we believe more accurately reflects the realities of designing such attacks in practice.

\section*{Acknowledgments}
The submitted manuscript has been created by UChicago Argonne, LLC, Operator of Argonne National Laboratory (``Argonne”).
Argonne, a U.S. Department of Energy Office of Science laboratory, is operated under Contract No. DE-AC02-06CH11357. 
The U.S. Government retains for itself, and others acting on its behalf, a paid-up nonexclusive, irrevocable worldwide license in said article to reproduce, prepare derivative works, distribute copies to the public, and perform publicly and display publicly, by or on behalf of the Government. 
The Department of Energy will provide public access to these results of federally sponsored research in accordance with the DOE Public Access Plan.

\bibliographystyle{unsrtnat}
\bibliography{references}

\end{document}